\documentclass[12pt]{article}
\usepackage{lmodern}
\usepackage{lmodern}
\usepackage[T1]{fontenc}
\usepackage[utf8]{inputenc}
\usepackage{geometry}
\usepackage{color}
\usepackage{mathrsfs}
\usepackage{url}
\usepackage{graphicx}
\usepackage{setspace}
\PassOptionsToPackage{version=3}{mhchem}
\usepackage{mhchem}
\usepackage[numbers]{natbib}
\usepackage[unicode=true,
 bookmarks=true,bookmarksnumbered=false,bookmarksopen=false,
 breaklinks=true,pdfborder={0 0 0},pdfborderstyle={},backref=false,colorlinks=true]
 {hyperref}
\hypersetup{
 citecolor=blue,linkcolor=blue,urlcolor=blue}

\makeatletter

\providecommand{\tabularnewline}{\\}

\newcommand{\lyxaddress}[1]{
	\par {\raggedright #1
	\vspace{1.4em}
	\noindent\par}
}

\usepackage{datetime}
\usepackage{refstyle}
\usepackage{longtable}
\usepackage{url}
\usepackage[title,toc,page,header]{appendix} 
\PassOptionsToPackage{version=3}{mhchem}

\makeatother

\begin{document}

\title{Impact of non-normal error distributions \\on the benchmarking and
ranking \\of Quantum Machine Learning models}

\author{Pascal Pernot$^{1}$, Bing Huang$^{2}$ and Andreas Savin$^{3}$}
\maketitle

\lyxaddress{$^{1}$ Institut de Chimie Physique, UMR8000, CNRS, Université Paris-Saclay,
91405 Orsay, France.\\
Contact: Pascal.Pernot@universite-paris-saclay.fr}

\lyxaddress{$^{2}$ Institute of Physical Chemistry and National Center for Computational
Design and Discovery of Novel Materials (MARVEL), Department of Chemistry,
University of Basel, Klingelbergstrasse 80, 4056 Basel, Switzerland.}

\lyxaddress{$^{3}$ Laboratoire de Chimie Théorique, CNRS and Sorbonne Universités, 75252 Paris, France.}

\begin{abstract}
\noindent Quantum machine learning models have been gaining significant
traction within atomistic simulation communities. Conventionally,
relative model performances are being assessed and compared using
learning curves (prediction error vs.~training set size). This article
illustrates the limitations of using the Mean Absolute Error (MAE)
for benchmarking, which is particularly relevant in the case of non-normal
error distributions. We analyze more specifically the prediction error
distribution of the kernel ridge regression with SLATM representation
and $L_{2}$ distance metric (KRR-SLATM-L2) for effective atomization
energies of QM7b molecules calculated at the level of theory CCSD(T)/cc-pVDZ.
Error distributions of HF and MP2 at the same basis set referenced
to CCSD(T) values were also assessed and compared to the KRR model.
We show that the true performance of the KRR-SLATM-L2 method over
the QM7b dataset is poorly assessed by the Mean Absolute Error, and
can be notably improved after adaptation of the learning set.
\end{abstract}

\newpage
\section{Introduction\label{sec:Introduction}}

For users to appreciate the accuracy of a computational chemistry
method, one should ideally provide them some statistics enabling to
estimate \emph{easily} the probability to get errors below a threshold
corresponding to their requirements. As usual for physical measurements,
the most simple statistic would be a prediction uncertainty attached
to a method $u(M)$ \citep{Pernot2015}. 

If the distribution of residual errors for a benchmark dataset is
zero-centered normal, statistics such as the mean absolute error (MAE),
the root mean squared deviation (RMSD) or the 95$^{th}$ quantile
of the absolute errors distribution ($Q_{95}$) are redundant and
can be used to infer $u(M)$: $u(M)\simeq\mathrm{RMSD}=\sqrt{\pi/2}\mathrm{MAE}\simeq0.5Q_{95}$.
If it is not normal, more information is necessary to provide the
user with probabilistic diagnostics \citep{Pernot2018}. It is therefore
primordial to test the normality of the distribution and, in case
of non-normality, to consider data transformations that might get
the distribution closer to normality. For instance, the use of intensive
properties \citep{Perdew2016,Pernot2018}, the choice between errors
or relative errors \citep{Lejaeghere2016}, and the correction of
trends \citep{Pernot2015} might have a notable impact.

The previous analysis has been applied to computational chemistry
methods, but to our knowledge, the error distributions of ML algorithms
have not be scrutinized for their ability to deliver a reliable prediction
uncertainty, and the general use of the MAE as a benchmark statistic
for ML methods \citep{Faber2017,Zaspel2019} has to be evaluated.
A problem arises notably when comparing methods with different error
distribution shapes, as MAE-based ranking might become arbitrary,
occulting important considerations about the risk of large errors
for some of the methods \citep{Pernot2018,Pernot2020a} .

This is the main topic of the present paper, where we analyze the
prediction errors for effective atomization energies of QM7b molecules
calculated at the level of theory CCSD(T)/cc-pVDZ by the kernel ridge
regression with CM and SLATM representations and $L_{2}$ distance
metric \citep{Zaspel2019}. The ML error distributions are compared
with the ones obtained from computational chemistry methods (HF and
MP2) on the same reference dataset. 

In the next section, we present the statistical tools we use to characterize
error sets. These are then applied to the HF, MP2, CM-L2 and SLATM-L2
error sets (Section\,\ref{sec:Application}). The strong non-normality
of the SLATM-L2 errors is then scrutinized and the systems having
large errors are analyzed as outliers. The impact on the prediction
errors distribution of including these outliers in the learning set
is evaluated. The main findings are discussed in Section\,\ref{sec:Discussion}.

\section{Statistical methods\label{sec:Statistical-methods}}

Statistical benchmarking of a method $M$ is based on the estimation
of errors ($E_{M}=\left\{ e_{M,i}\right\} _{i=1}^{N}$) for a set
of $N$ calculated ($C_{M}=\left\{ c_{M,i}\right\} _{i=1}^{N}$) and
reference data ($R=\left\{ r_{i}\right\} _{i=1}^{N}$), where
\begin{equation}
e_{M,i}=r_{i}-c_{M,i}
\end{equation}
In the present study, reference data are calculated by a quantum chemistry
method, and uncertainty on the calculated and reference values are
considered to be negligible before the errors.

\subsection{Normality assesment\label{subsec:(Non-)-Normality-assesment}}

Reliable use of statistical tests of normality require typically at
least $N=100$ points \citep{Razali2011,Klauenberg2019a}. For a given
sample size, the Shapiro-Wilk $W$ statistics has been shown to have
good properties \citep{Razali2011}, and it is used in this study.
The values of $W$ range between 0 and 1, and values of $W\simeq1$
are in favor of the normality of the sample. If $W$ lies below a
critical value $W_{c}$ depending on the sample size and the chosen
level of type I errors $\alpha$ (typically 0.05), the normality hypothesis
cannot be assumed to hold \citep{Klauenberg2019a}.

It might also be useful to assess normality by visual tools: normal
quantile-quantile plots (QQ-plots) \citep{Lejaeghere2016}, where
the quantiles of the scaled and centered errors sample is plotted
against the theoretical quantiles of a standard normal distribution
(in the normal case, all points should lie over the unit line); or
comparison of the histogram of errors with a gaussian curve having
the same mean, estimated by the mean signed error (MSE), and same
standard deviation, estimated by the RMSD. 

Two other statistics are helpful in characterizing the departure from
normality. The skewness (Skew), or third standardized moment of the
distribution, quantifies its asymmetry (Skew\,=\,0 for a symmetric
distribution). The kurtosis (Kurt), or fourth standardized moment,
quantifies the concentration of data in the tails of the distribution.
Kurtosis of a normal distribution is equal to 3; distributions with
excess kurtosis (Kurt\,>\,3) are called \emph{leptokurtic}; those
with Kurt\,<\,3 are named \emph{platykurtic}. 

For error distributions which are non symmetric (Skew\,$\ne0$),
quantifying the accuracy by a single dispersion-related statistic
is not reliable, and one should provide probability intervals or accept
to loose information on the sign and use a statistic based on absolute errors,
such as $Q_{95}$ presented below. 

For platykurtic and leptokurtic error distribution, the normal probabilistic
interpretation of dispersion statistics such as the RMSD is lost,\footnote{When Kurt$\ne3$, the interval $\mu\pm\sigma$ built on the mean $\mu$
(MSE) and standard deviation $\sigma$ (RMSD) is not a 67\,\% probability
interval.} and one should rely on dispersion measures providing explicit probability
coverage, such as $u_{95}$ recommended in the thermochemistry literature
\citep{Ruscic2014}. $u_{95}$ is an enlarged uncertainty which enables
to define a symmetric 95\,\% probability interval around the estimated
value. If the distribution is also skewed, one cannot rely on a symmetric
interval based on $u_{95}$. A major problem with leptokurtic error
distributions is that the excess of data in the tails is related to
a risk of large prediction errors, that cannot be appreciated from
the usual MAE or RMSD statistics. A solution is again to use $Q_{95}$
(Section\,\ref{subsec:Empirical-Cumulative-Distributio}).

For ranking studies, problems might occur when comparing distribution
with different shapes, which might lead to conflict between different
ranking statistics \citep{Pernot2018}. A striking example is provided
below (Section\,\ref{sec:Application}).

\subsection{Empirical Cumulative Distribution Function and related statistics\label{subsec:Empirical-Cumulative-Distributio}}

Pernot and Savin \citep{Pernot2018} have shown that the non-normality
of error distributions prevents the probabilistic interpretation of
usual benchmark statistics. However, direct probabilistic information
can be extracted from the empirical cumulative distribution function
(ECDF) of absolute errors $\epsilon_{i}=|e_{i}|$
\begin{equation}
C(\eta)=\frac{1}{N}\sum_{i=1}^{N}\mathbf{1_{\epsilon_{i}\le\eta}}
\end{equation}
where $\eta$ is a threshold value to be chosen, and $\mathbf{1}_{x}$
is the indicator function, with value 1 if $x$ is true and 0 otherwise. 

Two statistics based on the ECDF have been proposed as fulfilling
the needs of end-users to choose a method suited to their purpose:
\begin{itemize}
\item $C(\eta)$ is the probability that the absolute errors lie below a
chosen threshold $\eta$, such as the ``chemical accuracy'' \citep{Pernot2018}.
A user can define his own needs and accordingly pick methods with
a high value of $C(\eta)$. 
\item $Q_{95}$, the 95$^{th}$ percentile of the absolute error distribution,
gives the amplitude of errors that there is a 5\% probability to exceed
\citep{Pernot2018}. The choice of this specific percentile was based
on several considerations. When the error distribution is normal,
$Q_{95}$ is identical to the enlarged uncertainty $u_{95}$ recommended
for thermochemical calculations and data \citep{Ruscic2014}. Besides,
$Q_{95}$ is much less sensitive to outliers than the maximal error,
or even $Q_{99}$ for small datasets. Thakkar \emph{et al.} \citep{Thakkar2015}
proposed a similar statistic ($P_{90}$) based on the 90$^{th}$ percentile.
\end{itemize}
As mentioned previously, the use of such probabilistic scores for
non-normal distributions relieves the ambiguity attached to the MAE
and RMSD. For instance, in all the cases of error sets that Pernot
and Savin \citep{Pernot2018,Pernot2020a} have studied previously,
the probability for an absolute error to be larger than the MAE lied
between 0.2 and 0.45 (it should be about 0.42 for a zero-centered
normal distribution). The small values are typically associated with
skewed and leptokurtic distributions. Several cases have also been
observed where two methods have similar MAE values and very different
probabilities of large errors, due to their different error distributions.

This non-ambiguity is essential if one wants to assess a risk of large
prediction errors. Note that the transition from descriptive statistics
to predictive statistics for risk assesment requires further constraints
on the reference dataset and error distributions. Ideally, the reference
dataset has to be representative enough to cover future prediction
cases and there should be no notable trend in the errors with respect
to the calculated values by the method of interest (in the statistical
prediction framework, the predictor variable is the calculated value
of a property, from which one wants to infer a credible interval for
its true value). 

Several statistical trend correction methods have been proposed in
the computational chemistry literature, from the simple scaling of
the calculated values \citep{Scott1996,Pernot2011}, or linear corrections
\citep{Lejaeghere2014,Lejaeghere2014a,Pernot2015,Lejaeghere2016,Proppe2017},
to more complex, ML-based corrections, such as $\Delta$-ML \citep{Ramakrishnan2015,Ward2019,Zaspel2019}
or Gaussian Processes \citep{Proppe2019a}. 

\subsection{Outliers identification\label{subsec:Outliers-identification}}

There is no unique method to identify outliers for a non-normal distribution.
One might, for instance, use visual tools, such as QQ-plots \citep{Lejaeghere2016},
or automatic selection tools, such as selecting points for which the
absolute error is larger than the 95-th percentile ($Q_{95}$), or
another percentile corresponding to a predefined error threshold.

In cases where the errors distribution seems heterogeneous, one can
attempt to analyze it as a mixture of normal distributions. One might
for instance use a bi-normal fit 
\begin{equation}
\mathscr{E}(x)=\sum_{i=1}^{2}w_{i}\mathcal{N}(x;\mu_{i},\sigma_{i})\label{eq:binorfit}
\end{equation}
where $\mathcal{N}(x;\mu,\sigma)$ is a normal distribution of $x$
with mean value $\mu$ and standard deviation $\sigma$. Outliers
are then defined as the points which lie outside of a $\mu_{1}\pm n\times\sigma_{1}$
interval, where $\mu_{1}$ and $\sigma_{1}$ are the parameters of
the most concentrated component. The enlargement factor $n$ should
be chosen large enough to exclude as much points from this component
as possible, but small enough to capture enough points of the wider
component to enable tests of chemical or physical hypotheses.

\subsection{Implementation}

All calculations have been made in the \texttt{R} language \citep{CiteR},
using several packages, notably for bootstrap \texttt{(boot} \citep{R-boot}),
skewness and kurtosis (\texttt{moments} \citep{R-moments}) and mixture
analysis (\texttt{mixtools} \citep{R-mixtools,Benaglia2009}). Bootstrap
estimates are based on 1000 draws. The \texttt{R} implementation of
the Shapiro-Wilk test (function \texttt{shapiro.test}) has a limit
of dataset size at 5000 \citep{Razali2011}. For the large datasets
used in the application, 5000 points are randomly chosen to evaluate
$W$.

The code and datasets to reproduce the tables and figures of this
article are available at Zenodo (\url{https://doi.org/10.5281/zenodo.3733272}).
The datasets can also be analyzed with the ErrView code (\url{https://doi.org/10.5281/zenodo.3628489}),
or its web interface (\url{http://upsa.shinyapps.io/ErrView}).

\section{Application \label{sec:Application}}

The data are issued from the study by Zaspel \emph{et al.} \citep{Zaspel2019}.
The effective atomization energies ($E^{*}$) for the QM7b dataset
\citep{Montavon2013}, for 7211 molecules up to 7 heavy atoms (C,
N, O, S or Cl) are available for several basis sets (STO-3g, 6-31g,
and cc-pVDZ), three quantum chemistry methods (HF, MP2 and CCSD(T))
and four machine learning algorithms (CM-L1, CM-L2, SLATM-L1 and SLATM-L2).
The ML methods have been trained over a random sample of 1000 CCSD(T)
energies (learning set), and the test set contains the prediction
errors for the 6211 remaining systems \citep{Zaspel2019}. We consider
here the results for the cc-pVDZ basis set and the CM-L2 and SLATM-L2
methods. Besides, the errors for HF and MP2 have been estimated on
the same reference data set as for the ML methods. They provide an
interesting contrast in terms of error distribution. 
\begin{table}[t]
\noindent \begin{centering}
\begin{tabular}{lr@{\extracolsep{0pt}.}lr@{\extracolsep{0pt}.}lr@{\extracolsep{0pt}.}lr@{\extracolsep{0pt}.}lr@{\extracolsep{0pt}.}lr@{\extracolsep{0pt}.}lr@{\extracolsep{0pt}.}lr@{\extracolsep{0pt}.}l}
\hline 
{\scriptsize{}Methods } & \multicolumn{2}{c}{{\scriptsize{}MAE }} & \multicolumn{2}{c}{{\scriptsize{}MSE }} & \multicolumn{2}{c}{{\scriptsize{}RMSD }} & \multicolumn{2}{c}{{\scriptsize{}$Q_{95}$}} & \multicolumn{2}{c}{{\scriptsize{}$C(\eta)$}} & \multicolumn{2}{c}{{\scriptsize{}Skew }} & \multicolumn{2}{c}{{\scriptsize{}Kurt }} & \multicolumn{2}{c}{{\scriptsize{}$W$}}\tabularnewline
\cline{2-11} \cline{16-17} 
 & \multicolumn{8}{c}{{\scriptsize{}(kcal/mol) }} & \multicolumn{2}{c}{{\scriptsize{}$\eta=1$\,kcal/mol}} & \multicolumn{2}{c}{} & \multicolumn{2}{c}{{\scriptsize{} }} & \multicolumn{2}{c}{{\scriptsize{}$W_{c}=0.9993$}}\tabularnewline
\hline 
{\scriptsize{}HF } & {\scriptsize{}2}&{\scriptsize{}38(3) } & {\scriptsize{}0}&{\scriptsize{}06(4) } & {\scriptsize{}3}&{\scriptsize{}13(4) } & {\scriptsize{}6}&{\scriptsize{}1(1) } & {\scriptsize{}0}&{\scriptsize{}273(6) } & {\scriptsize{}-0}&{\scriptsize{}69(7) } & {\scriptsize{}5}&{\scriptsize{}1(3) } & {\scriptsize{}0}&{\scriptsize{}970(4) }\tabularnewline
{\scriptsize{}MP2 } & {\scriptsize{}1}&{\scriptsize{}31(1) } & {\scriptsize{}0}&{\scriptsize{}00(2) } & {\scriptsize{}1}&{\scriptsize{}67(2) } & {\scriptsize{}3}&{\scriptsize{}35(5) } & {\scriptsize{}0}&{\scriptsize{}461(6) } & {\scriptsize{}-0}&{\scriptsize{}07(3) } & {\scriptsize{}3}&{\scriptsize{}33(6) } & {\scriptsize{}0}&{\scriptsize{}9961(8) }\tabularnewline
{\scriptsize{}CM-L2 } & {\scriptsize{}10}&{\scriptsize{}1(1) } & {\scriptsize{}-0}&{\scriptsize{}0(2) } & {\scriptsize{}13}&{\scriptsize{}2(2) } & {\scriptsize{}26}&{\scriptsize{}2(4) } & {\scriptsize{}0}&{\scriptsize{}071(3) } & {\scriptsize{}0}&{\scriptsize{}05(7) } & {\scriptsize{}4}&{\scriptsize{}3(3) } & {\scriptsize{}0}&{\scriptsize{}990(3) }\tabularnewline
{\scriptsize{}SLATM-L2 } & {\scriptsize{}1}&{\scriptsize{}26(3) } & {\scriptsize{}0}&{\scriptsize{}13(3) } & {\scriptsize{}2}&{\scriptsize{}44(8) } & {\scriptsize{}4}&{\scriptsize{}7(1) } & {\scriptsize{}0}&{\scriptsize{}651(6) } & {\scriptsize{}0}&{\scriptsize{}8(5) } & {\scriptsize{}29}&{\scriptsize{}(3) } & {\scriptsize{}0}&{\scriptsize{}70(2) }\tabularnewline
\hline 
{\scriptsize{}lc-HF } & {\scriptsize{}2}&{\scriptsize{}29(3) } & {\scriptsize{}0}&{\scriptsize{}0 } & {\scriptsize{}2}&{\scriptsize{}97(4) } & {\scriptsize{}5}&{\scriptsize{}87(8) } & {\scriptsize{}0}&{\scriptsize{}285(6) } & {\scriptsize{}-0}&{\scriptsize{}46(6) } & {\scriptsize{}4}&{\scriptsize{}3(2) } & {\scriptsize{}0}&{\scriptsize{}985(3) }\tabularnewline
{\scriptsize{}lc-MP2 } & {\scriptsize{}1}&{\scriptsize{}28(1) } & {\scriptsize{}0}&{\scriptsize{}0 } & {\scriptsize{}1}&{\scriptsize{}61(1) } & {\scriptsize{}3}&{\scriptsize{}18(5) } & {\scriptsize{}0}&{\scriptsize{}470(6) } & {\scriptsize{}0}&{\scriptsize{}09(3) } & {\scriptsize{}3}&{\scriptsize{}14(6) } & {\scriptsize{}0}&{\scriptsize{}9983(5) }\tabularnewline
{\scriptsize{}lc-CM-L2} & {\scriptsize{}10}&{\scriptsize{}1(1) } & {\scriptsize{}0}&{\scriptsize{}0} & {\scriptsize{}13}&{\scriptsize{}1(2) } & {\scriptsize{}26}&{\scriptsize{}4(4) } & {\scriptsize{}0}&{\scriptsize{}071(3)} & {\scriptsize{}0}&{\scriptsize{}09(7) } & {\scriptsize{}4}&{\scriptsize{}4(3) } & {\scriptsize{}0}&{\scriptsize{}990(3)}\tabularnewline
{\scriptsize{}lc-SLATM-L2 } & {\scriptsize{}1}&{\scriptsize{}27(3) } & {\scriptsize{}0}&{\scriptsize{}0} & {\scriptsize{}2}&{\scriptsize{}44(8) } & {\scriptsize{}4}&{\scriptsize{}7(1) } & {\scriptsize{}0}&{\scriptsize{}650(6) } & {\scriptsize{}0}&{\scriptsize{}8(5) } & {\scriptsize{}28}&{\scriptsize{}(3) } & {\scriptsize{}0}&{\scriptsize{}71(2) }\tabularnewline
\hline 
{\scriptsize{}SLATM-L2(-o) } & {\scriptsize{}0}&{\scriptsize{}85(1) } & {\scriptsize{}0}&{\scriptsize{}06(2) } & {\scriptsize{}1}&{\scriptsize{}20(2) } & {\scriptsize{}2}&{\scriptsize{}71(4) } & {\scriptsize{}0}&{\scriptsize{}690(6) } & {\scriptsize{}-0}&{\scriptsize{}04(5) } & {\scriptsize{}4}&{\scriptsize{}8(1) } & {\scriptsize{}0}&{\scriptsize{}968(3) }\tabularnewline
{\scriptsize{}SLATM-L2(+o) } & {\scriptsize{}0}&{\scriptsize{}83(1) } & {\scriptsize{}0}&{\scriptsize{}05(2) } & {\scriptsize{}1}&{\scriptsize{}20(2) } & {\scriptsize{}2}&{\scriptsize{}49(5) } & {\scriptsize{}0}&{\scriptsize{}713(6) } & {\scriptsize{}-0}&{\scriptsize{}1(1) } & {\scriptsize{}8}&{\scriptsize{}1(5) } & {\scriptsize{}0}&{\scriptsize{}940(7) }\tabularnewline
{\scriptsize{}SLATM-L2(+r) } & {\scriptsize{}0}&{\scriptsize{}95(2) } & {\scriptsize{}-0}&{\scriptsize{}01(2) } & {\scriptsize{}1}&{\scriptsize{}89(8) } & {\scriptsize{}3}&{\scriptsize{}15(8) } & {\scriptsize{}0}&{\scriptsize{}743(6) } & {\scriptsize{}0}&{\scriptsize{}4(10) } & {\scriptsize{}46}&{\scriptsize{}(7) } & {\scriptsize{}0}&{\scriptsize{}65(3) }\tabularnewline
\hline 
\end{tabular}
\par\end{centering}
\caption{\label{tab:MLStats} Error summary statistics on effective atomization
energies for the QM7b dataset. All calculations are done with the
cc-pvdz basis set, and the reference data are CCSD(T) values. Methods
with a 'lc-' prefix include linear correction. For the meaning of
the -o, +o and +r suffixes, see Sections\,\ref{subsec:Outliers-analysis}
and \ref{subsec:Expanding-the-learning}. MAE: mean absolute/unsigned
error; MSE: mean signed error; RMSD: root mean squared deviation;
$Q_{95}$: 95$^{th}$ percentile of the absolute errors distribution;
Skew, Kurt: skewness and kurtosis of the errors distribution; $C(\eta)$:
probability of absolute errors below $\eta=1$\,kcal/mol; $W$: Shapiro-Wilk
normality statistic. For a sample of size 5000, the critical value
is $W_{c}=0.9993$ at the 0.05 level. The normality hypothesis is
rejected if $W<W_{c}$. Standard uncertainties estimated by bootstrap
are reported in parenthetical notation.}
\end{table}

\subsection{Error statistics and distributions\label{subsec:Error-statistics} }

Summary statistics and their uncertainty have been estimated by bootstrap
\citep{Pernot2020}. They are reported in Table\,\ref{tab:MLStats}.
In this part, one concentrates on the unmodified methods, HF, MP2,
CM-L2, and SLATM-L2. Considering MAE alone, one might conclude that
SLATM-L2 produces slightly smaller errors than MP2 and a significant
improvement over all other methods. MSE tells us that there is no
important bias when compared to the MAE or RMSD (all distributions
are nearly centered on zero). It is striking that MP2 has a smaller
RMSD than SLATM-L2. As the RMSD is often taken as an estimator of
prediction uncertainty \citep{Pernot2015}, this contradicts the MAE
ranking. 
\begin{figure}[!p]
\noindent \centering{}%
\begin{tabular}{cc}
\includegraphics[width=0.27\textheight]{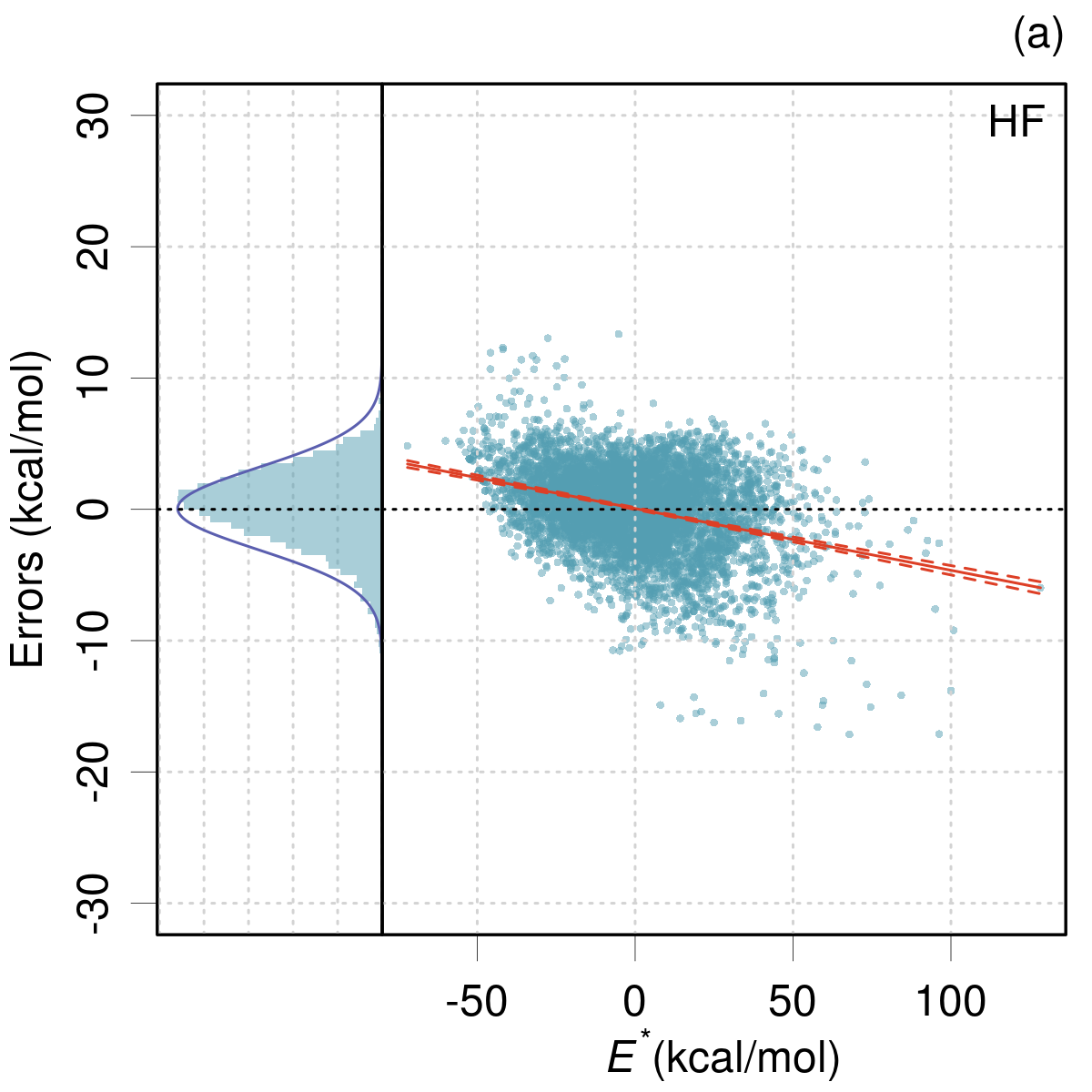} & \includegraphics[width=0.27\textheight]{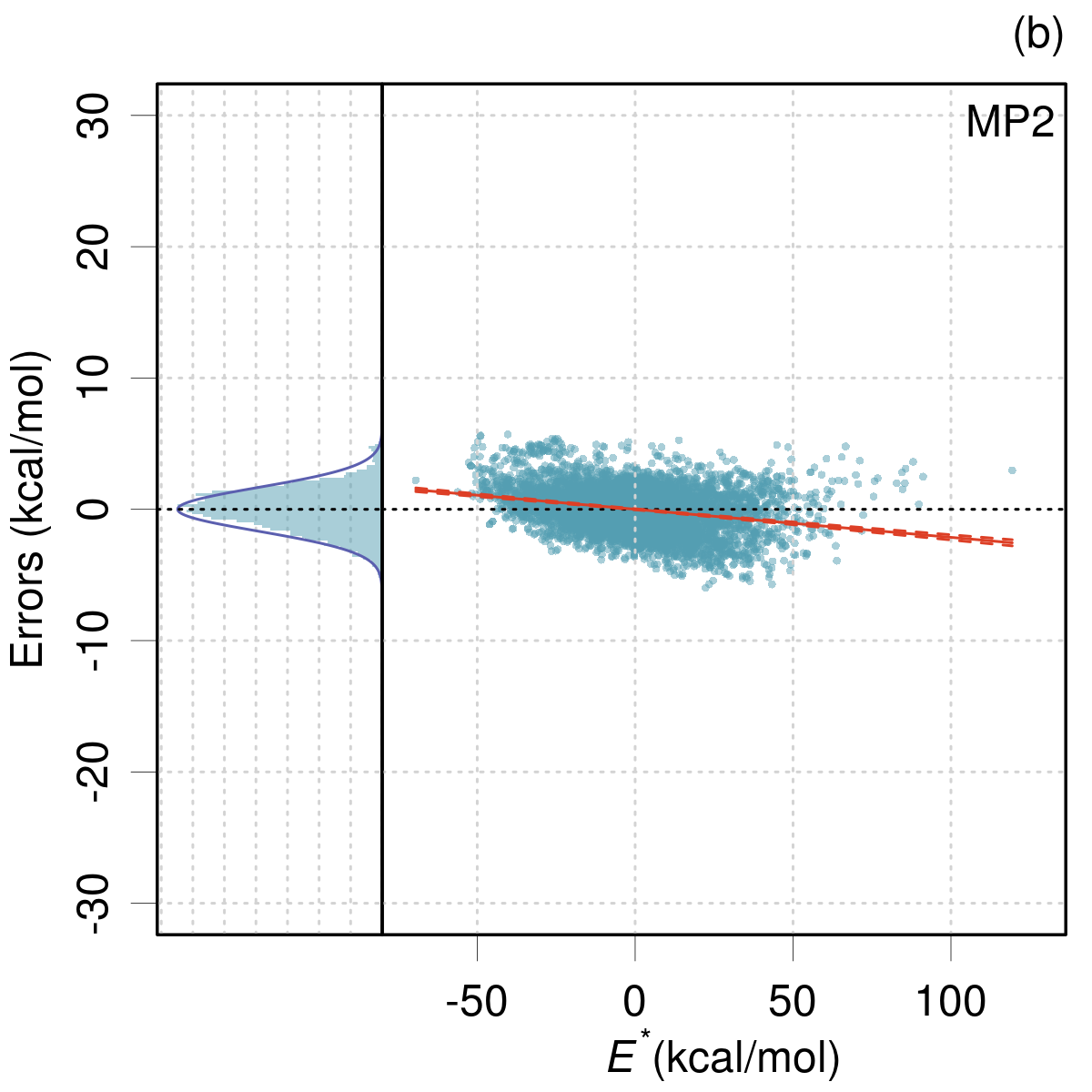}\tabularnewline
\includegraphics[width=0.27\textheight]{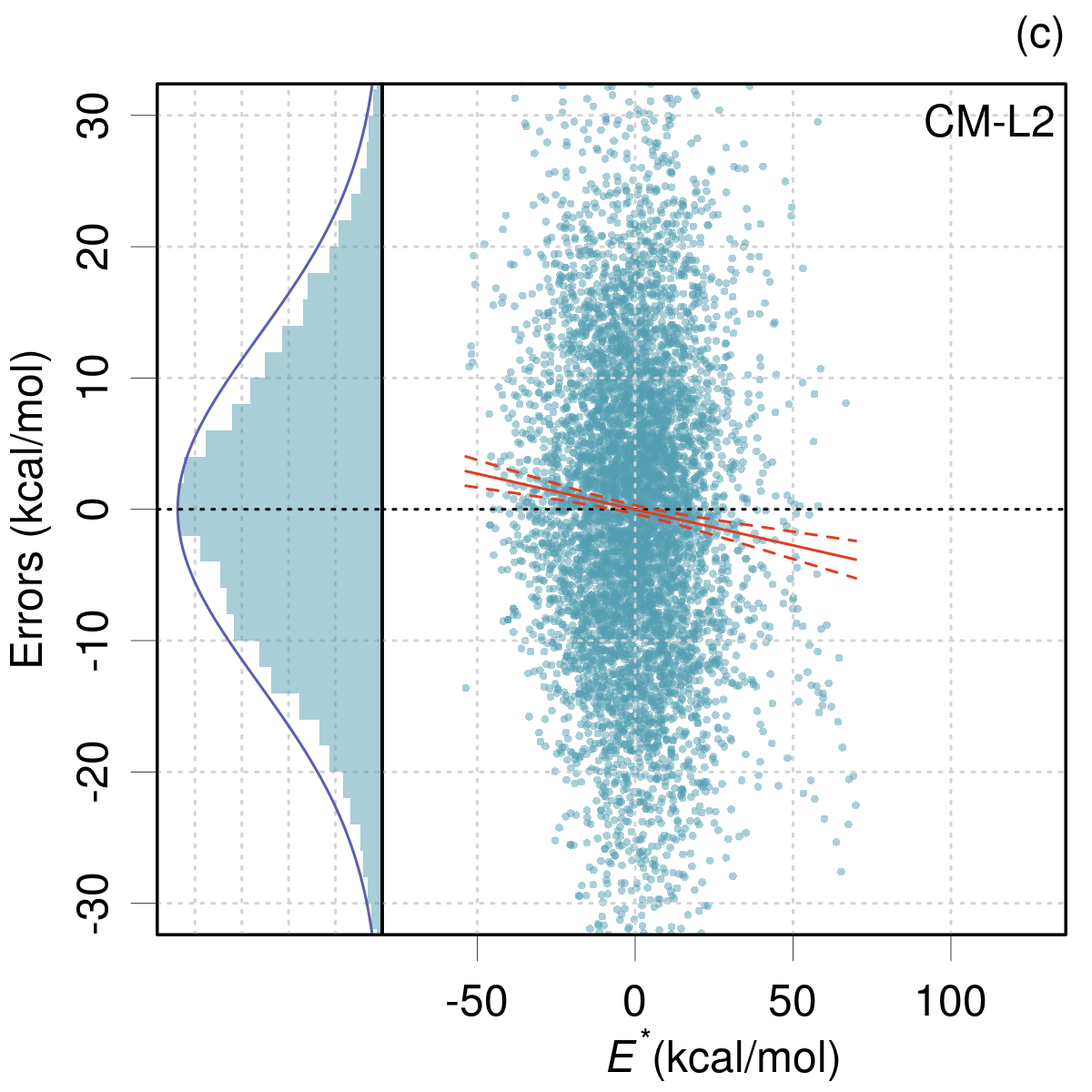} &  \includegraphics[width=0.27\textheight]{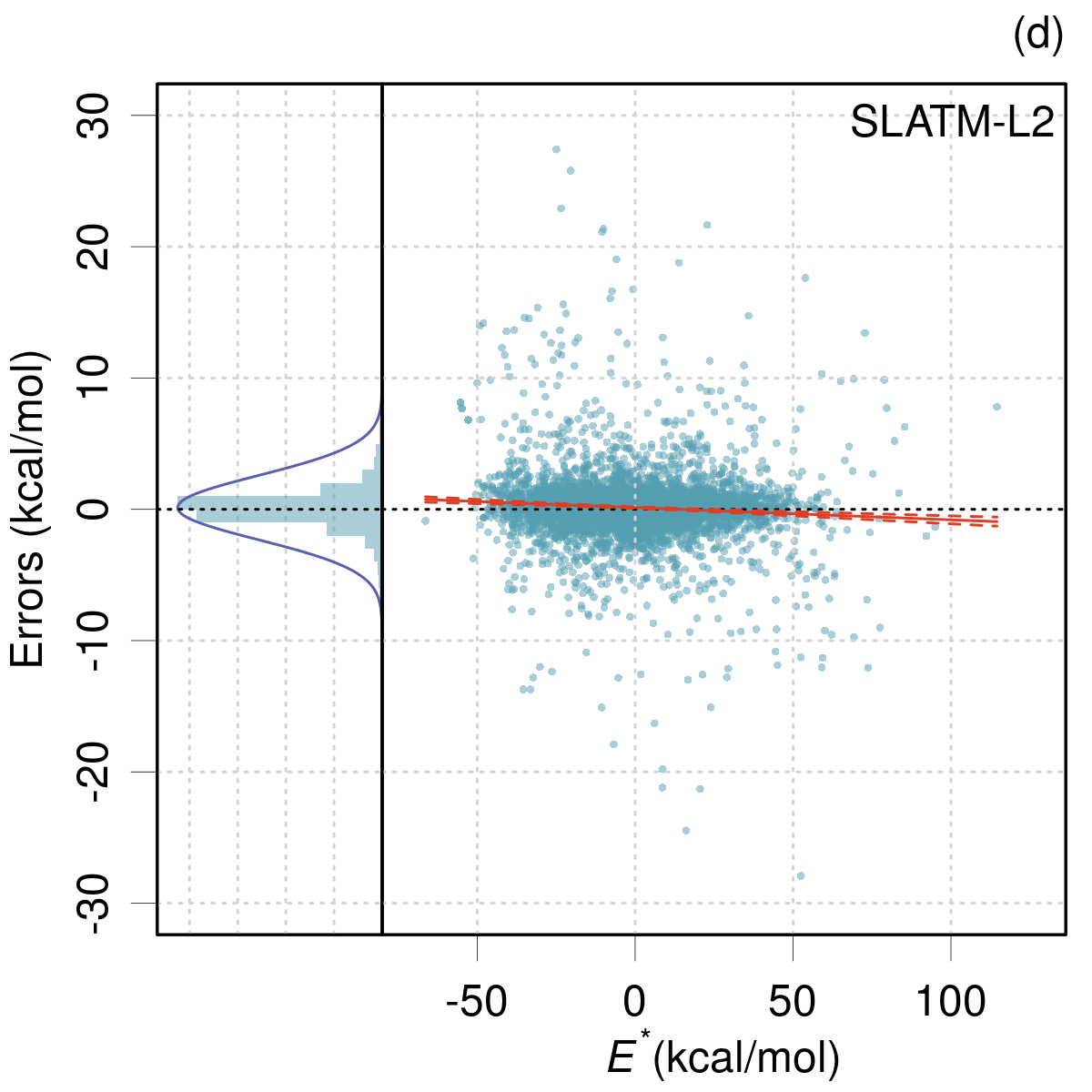}\tabularnewline
\includegraphics[width=0.27\textheight]{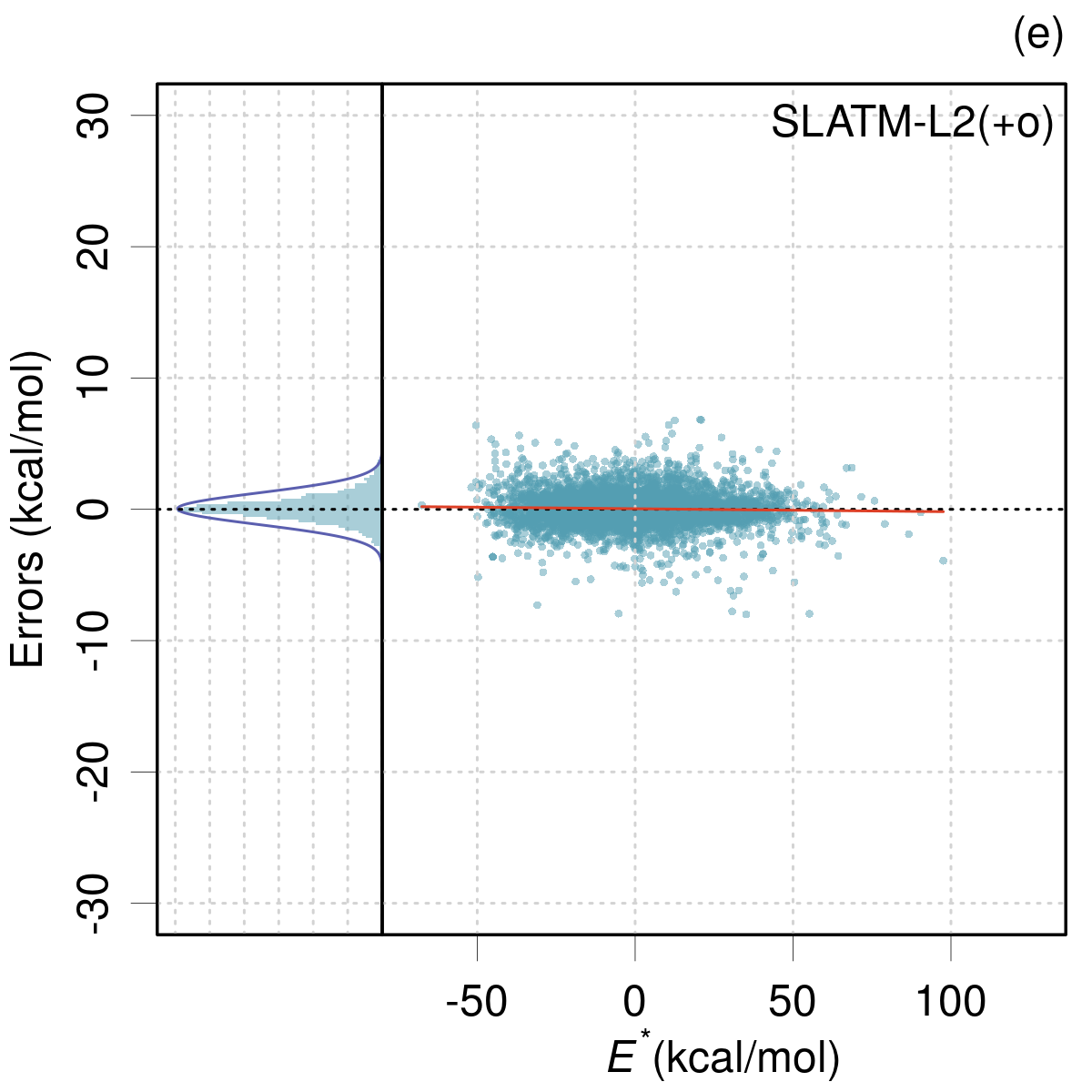}  &  \includegraphics[width=0.27\textheight]{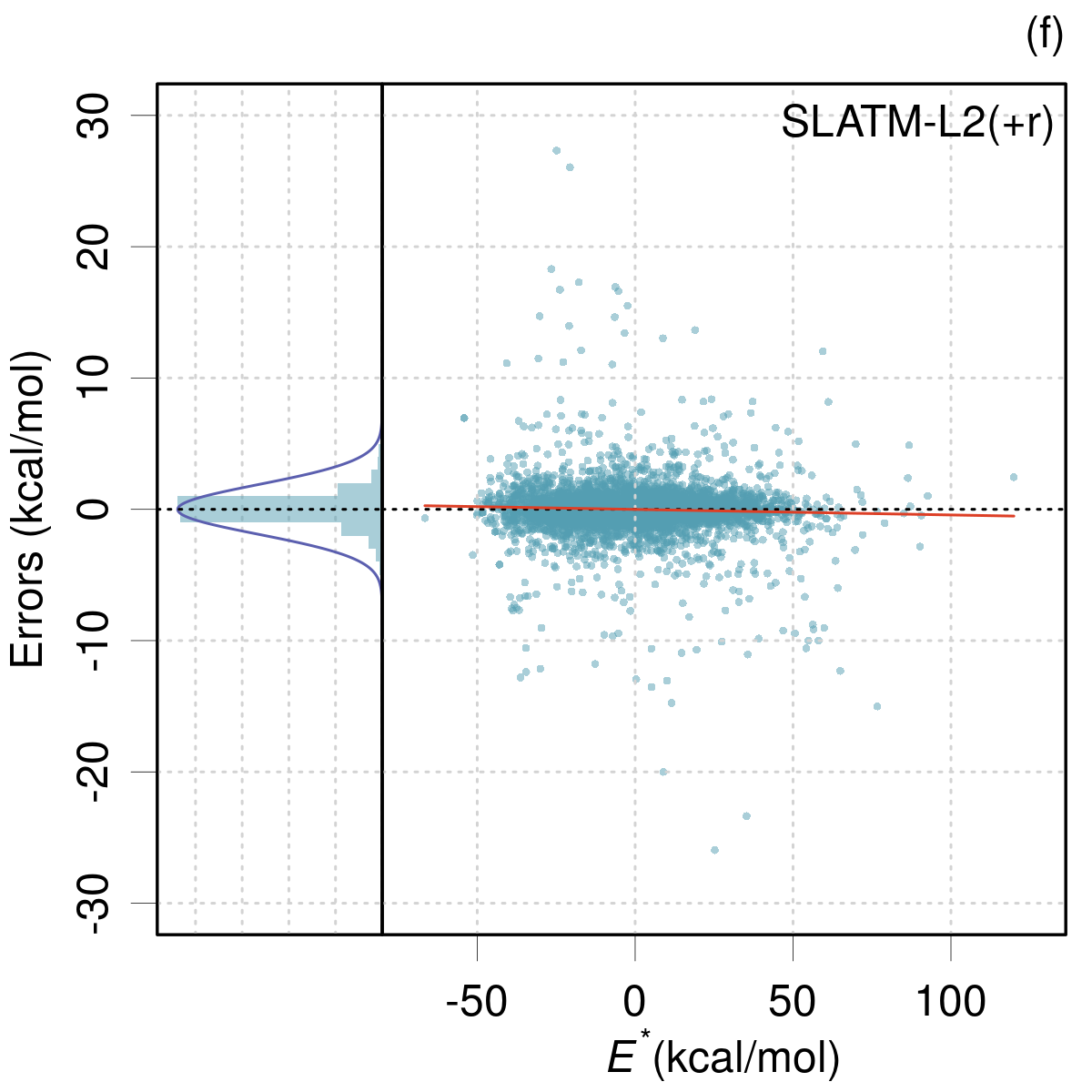}\tabularnewline
\end{tabular}\caption{\label{fig:ml1}Error distributions for the effective atomization
energies $E^{*}$ with respect to CCSD(T) of (a) HF, (b) MP2 (c) CM-L2,(d)
SLATM-L2, (e) SLATM-L2(+o) and (f) SLATM-L2(+r). In each sub-figure,
the left panel presents the histogram of the errors, with its fit
by a normal distribution, and the right panel a scatterplot of the
errors \emph{vs}. the calculated value by the corresponding method.
The red lines represent the linear trend with its 95\,\% confidence
interval, obtained by ordinary least squares linear regression. }
\end{figure}

This is also a strong clue that not all the error distributions are
normal. In the case of normal distributions, and considering that
MSE$\simeq0$, the RMSD should confirm the MAE ranking ($RMSD\simeq\sqrt{\pi/2}*MAE$
for a zero-centered normal distribution). Indeed, the Shapiro-Wilk
normality test rejects the normality of all the error sets, with a
critical value $W_{c}=0.9993$ for $N=5000$ and a type I error rate
of 0.05. However the $W$ statistic shows that MP2 is much closer
to normal ($W=0.9961$) than SLATM-L2 ($W=0.71$). This is confirmed
by the kurtosis, with a very large value for SLATM-L2 (29) whereas
for MP2 the value (3.33) is much closer to the value for a normal
distribution (3.0). HF and CM-L2 are also leptokurtic (Kurt\,>\,3),
but much less than SLATM-L2. 

The large kurtosis value indicates that the SLATM-L2 error distribution
presents tails much heavier than for a normal distribution. This excess
in large errors can be appreciated with $Q_{95}$, which informs us
that 5\,\% of the absolute errors are above 3.3\,kcal/mol for MP2 and
4.7\,kcal/mol for SLATM-L2. Despite a smaller MAE, one has thus a
non-negligible risk to get larger errors by SLATM-L2 than by MP2.
Note that in parallel with large errors, there is an excess in small
errors: we estimated that the probability to have absolute errors
smaller than the RMSD, $C(2.44)$, is 0.9 for SLATM-L2, when it should
be 0.67 for a normal distribution. 
\begin{figure}[!t]
\noindent \centering{}\includegraphics[width=0.45\textwidth]{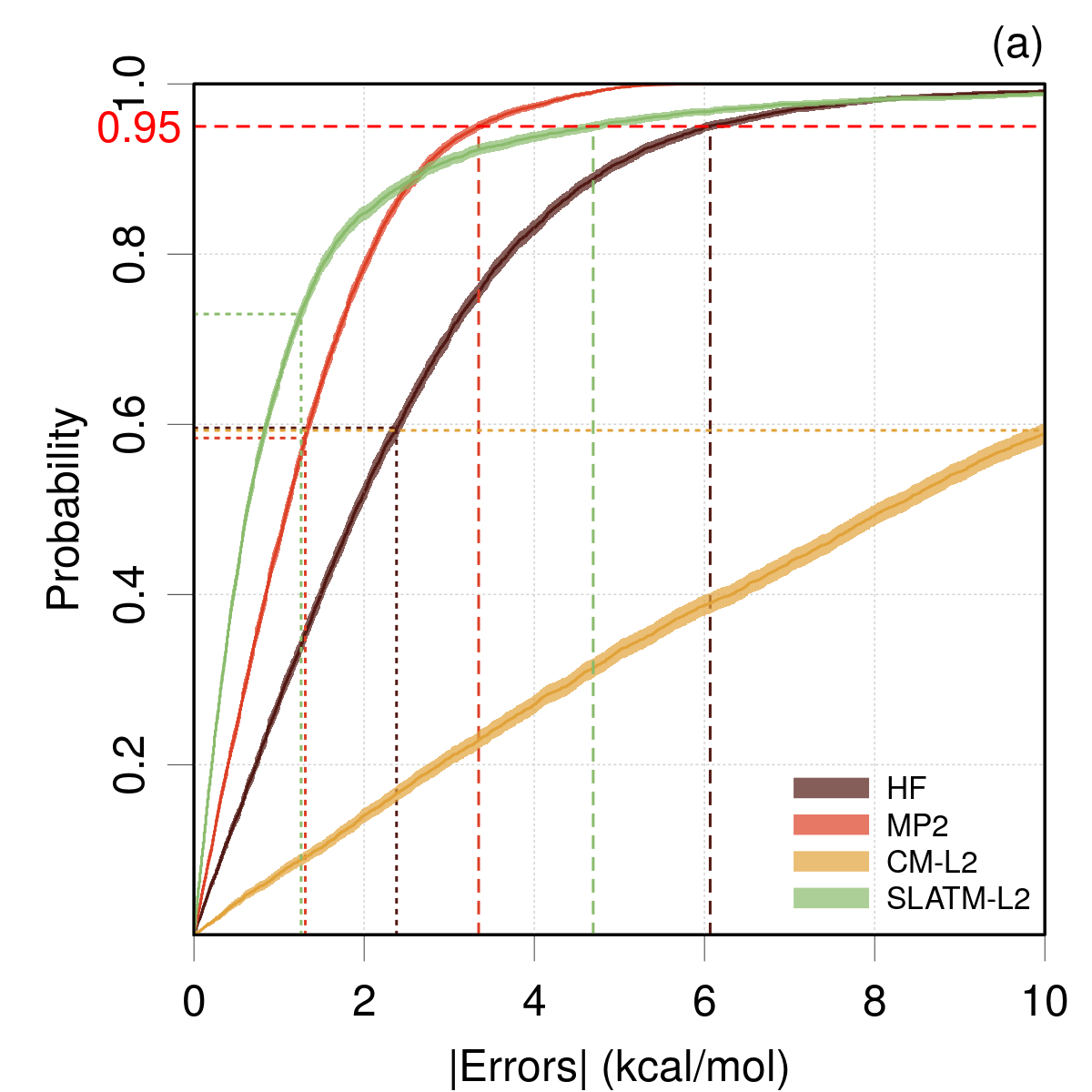}\includegraphics[width=0.45\textwidth]{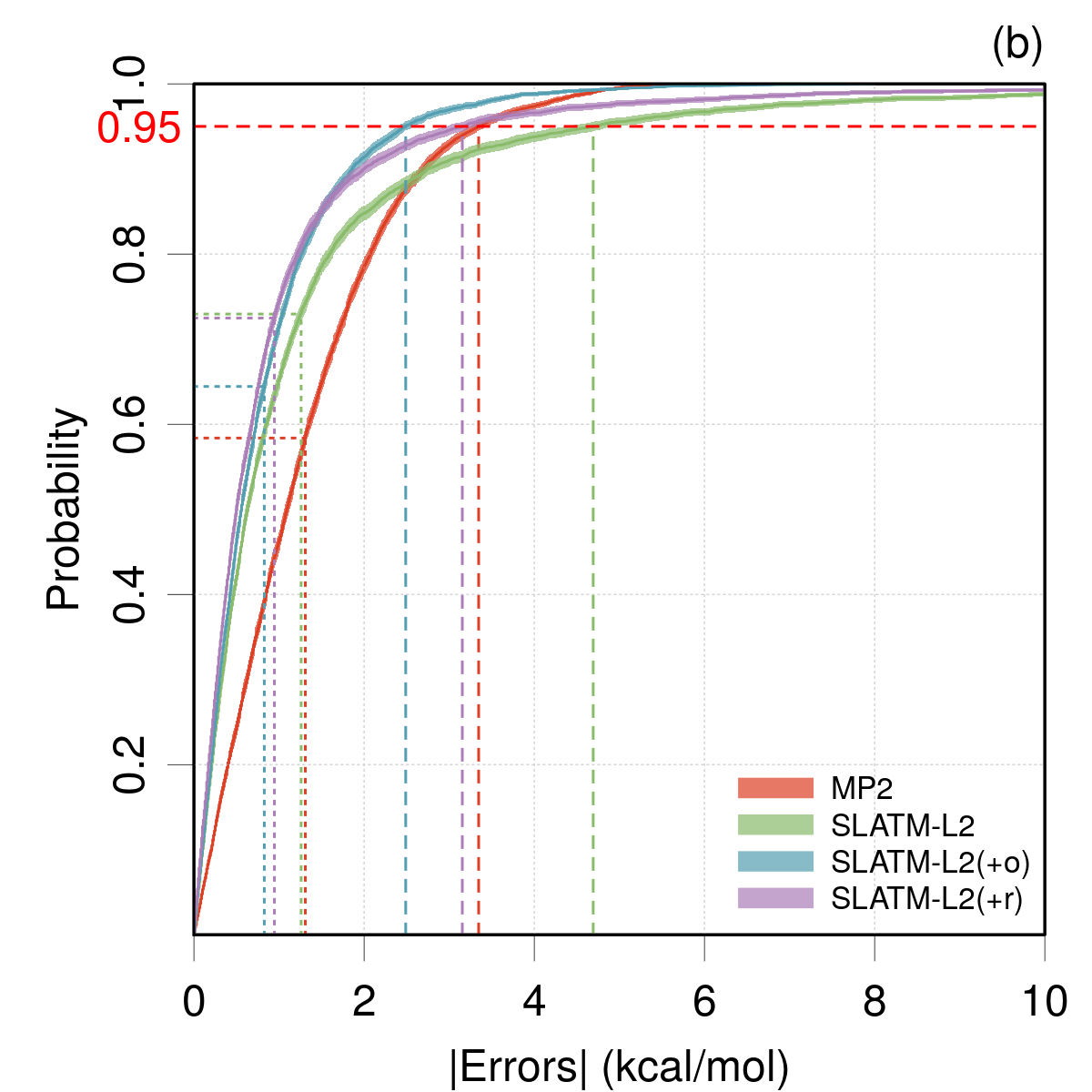}
\caption{\label{fig:ml1-1}ECDFs of the absolute errors the effective atomization
energies $E^{*}$ with respect to CCSD(T): (a) original methods; (b)
comparison with SLATM-L2 trained on a learning set augmented with
350 systems (+o: outliers of SLATM-L2; +r: random systems). The dashed
lines signal the $Q_{95}$ statistic for each set, and the dotted
lines the corresponding MAE.}
\end{figure}

\noindent In order to substantiate the interpretation of the error
statistics, it is important (1) to check that the error distributions
are not dominated by systematic errors (trends), and (2) to assess
their normality or lack thereof. This is essential if any quantification
of the predictive ability of the methods is of interest.

\paragraph{Trends. }

Plots of the error sets (Fig.\,\ref{fig:ml1}(a-d)) show that the
ML errors present weak (linear) trends, whereas HF and MP2 are more
affected. For comparison of their performances, the data sets have
been linearly corrected to remove the trends, according to Pernot
\emph{et al.}\citep{Pernot2015}. Summary statistics for corrected
methods (with a 'lc-' prefix) are reported in Table\,\ref{tab:MLStats},
showing a small impact on all statistics. The normality Shapiro-Wilk
statistic is improved for HF and MP2, and the error distribution for
lc-MP2 is practically normal. After linear correction, and considering
the small uncertainty on the linear correction parameters due to the
large size of the dataset, the prediction uncertainty can be estimated
by the RMSD \citep{Pernot2015}. This enables to assess the better
performance of MP2 in terms of prediction uncertainty. 

\paragraph{Normality.}

Comparison of the histograms with their gaussian fit in Fig.\,\ref{fig:ml1}(a-d)
shows clearly that the distributions for HF and MP2, albeit trended,
are nearly normal. Note that this is not a general trend of quantum
chemistry methods \citep{Pernot2018}. On the contrary, the error
distribution for SLATM-L2 is notably non-normal, with a narrow central
peak and wide tails. Because of the large number of points concentrated
near the center, the tails of the distribution are not visible on
the histogram in Fig.\,\ref{fig:ml1}(d). They are much more apparent
on the QQ-plot in Fig.\,\ref{fig:ml4}(a), which will be introduced
in Section\,\ref{subsec:Outliers-analysis}. 

\paragraph{ECDFs.}

Considering Fig.\,\ref{fig:ml1-1}, one see that the absolute errors
ECDFs for SLATM-L2 and MP2 intersect around 3\,kcal/mol. Below this
value, SLATM-L2 presents smaller absolute errors, with the opposite
above. It is clearly visible here why SLATM-L2 has a larger $Q_{95}$
than MP2. Indeed, one sees also on Figs.\,\ref{fig:ml1}\,(b,d)
that SLATM-L2 has a non-negligible set of errors much larger than
the largest MP2 errors. A similar effect occurs even between HF and
SLATM-L2, the latter presenting larger absolute errors above approximately
8\,kcal/mol.

\medskip{}

Computational time aside, the choice between MP2 and SLATM-L2 as a
predictor for CCSD(T) energies depends on the acceptance by the user
of a small percentage of large errors. SLATM-L2 would largely benefit
from a reduction of such large errors. In the following, we focus
on the SLATM-L2 error set to identify underlying features linked to
large prediction errors.

\subsection{Chemical analysis\label{subsec:Chemical-analysis}}

In order to get a chemical insight on the error trends in the present
dataset, the absolute errors have been analyzed as functions of several
compositional data, such as the number of individual species in the
composition (H, C, N, O, S, Cl) and the Double Bond Equivalent (DBE),
which estimates the degree of unsaturation (number of double bonds
and rings) \citep{Pellegrin1983} 
\begin{equation}
\mathrm{DBE}=n\mathrm{C}-\frac{n\mathrm{H}+n\mathrm{Cl}}{2}+\frac{n\mathrm{N}}{2}+1
\end{equation}
where $n\mathrm{X}$ is the number of $\mathrm{X}$ atoms in the molecule.
Note that in this formula divalent atoms do not contribute to the
DBE.

Sensitivity of the errors to the various variables has been estimated
by using rank correlation coefficients. The largest sensitivity has
been found with respect to the number of hydrogen atoms in the molecule
($n\mathrm{H}$) and the DBE. Note that both indicators are not independent
(DBE is anti-correlated with $n\mathrm{H}$), and that $n\mathrm{H}$
is strongly correlated with the number of atoms in the molecule. 
\begin{figure}[!t]
\noindent \centering{}%
\begin{tabular}{cc}
\includegraphics[width=0.45\textwidth]{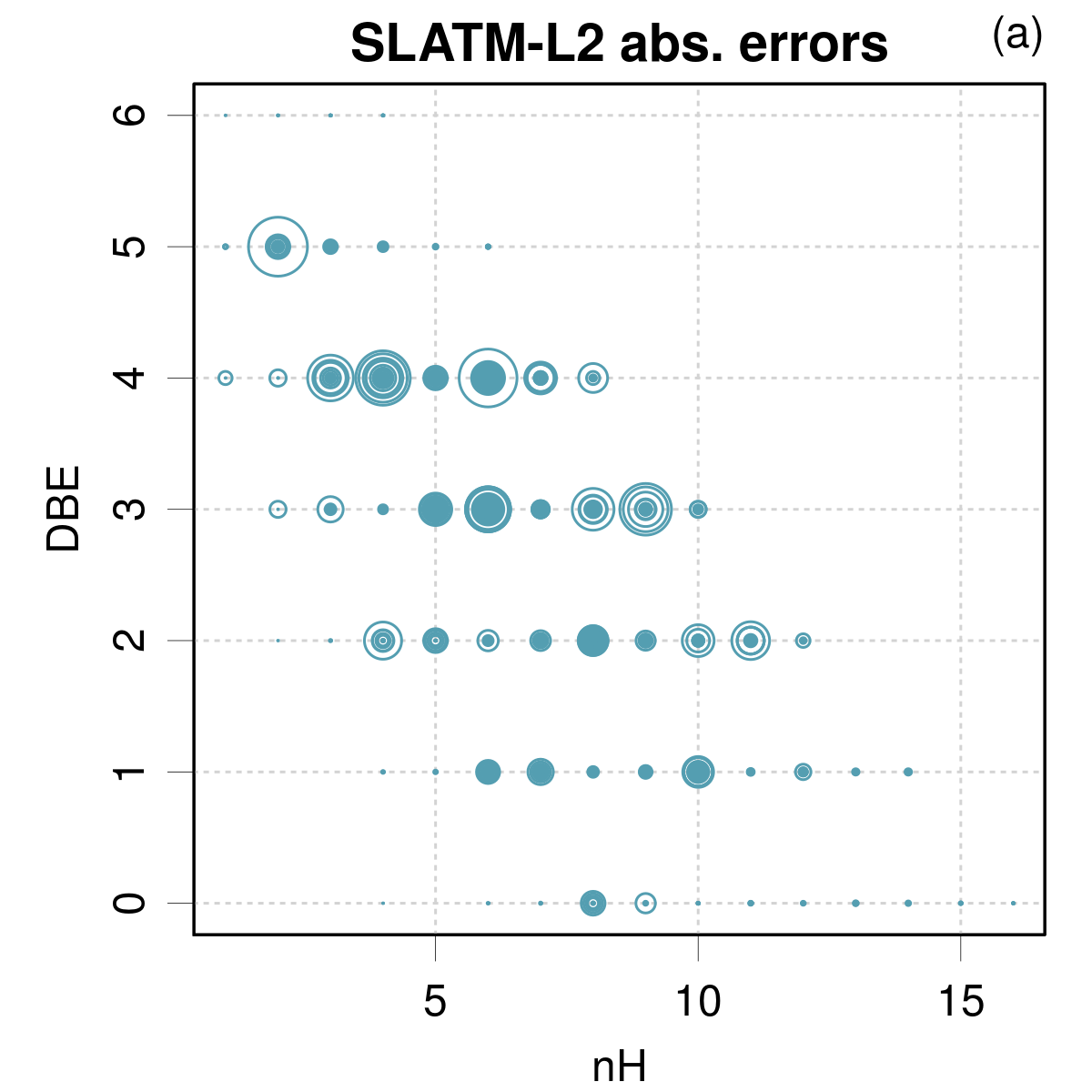}  & \includegraphics[width=0.45\textwidth]{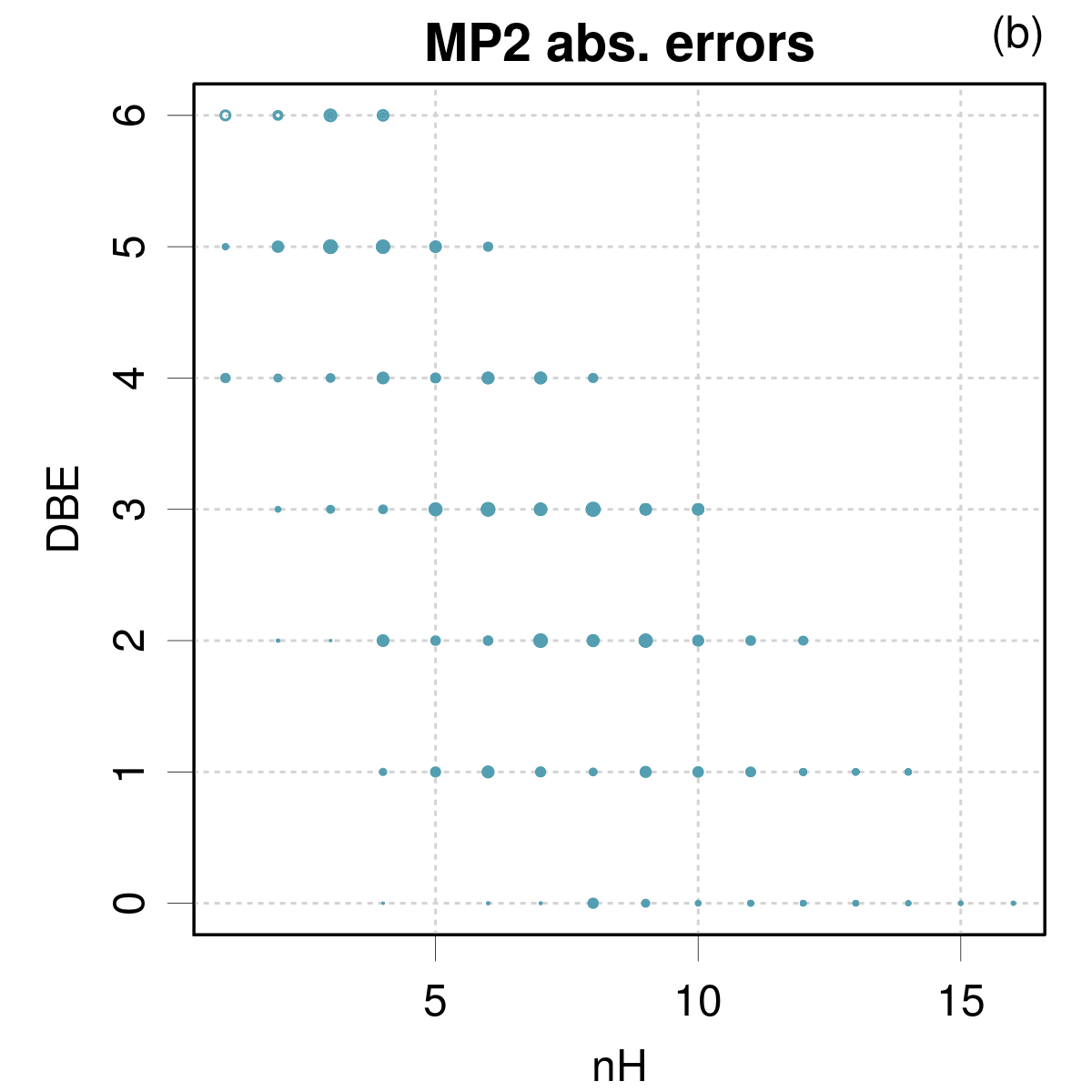} \tabularnewline
\includegraphics[width=0.45\textwidth]{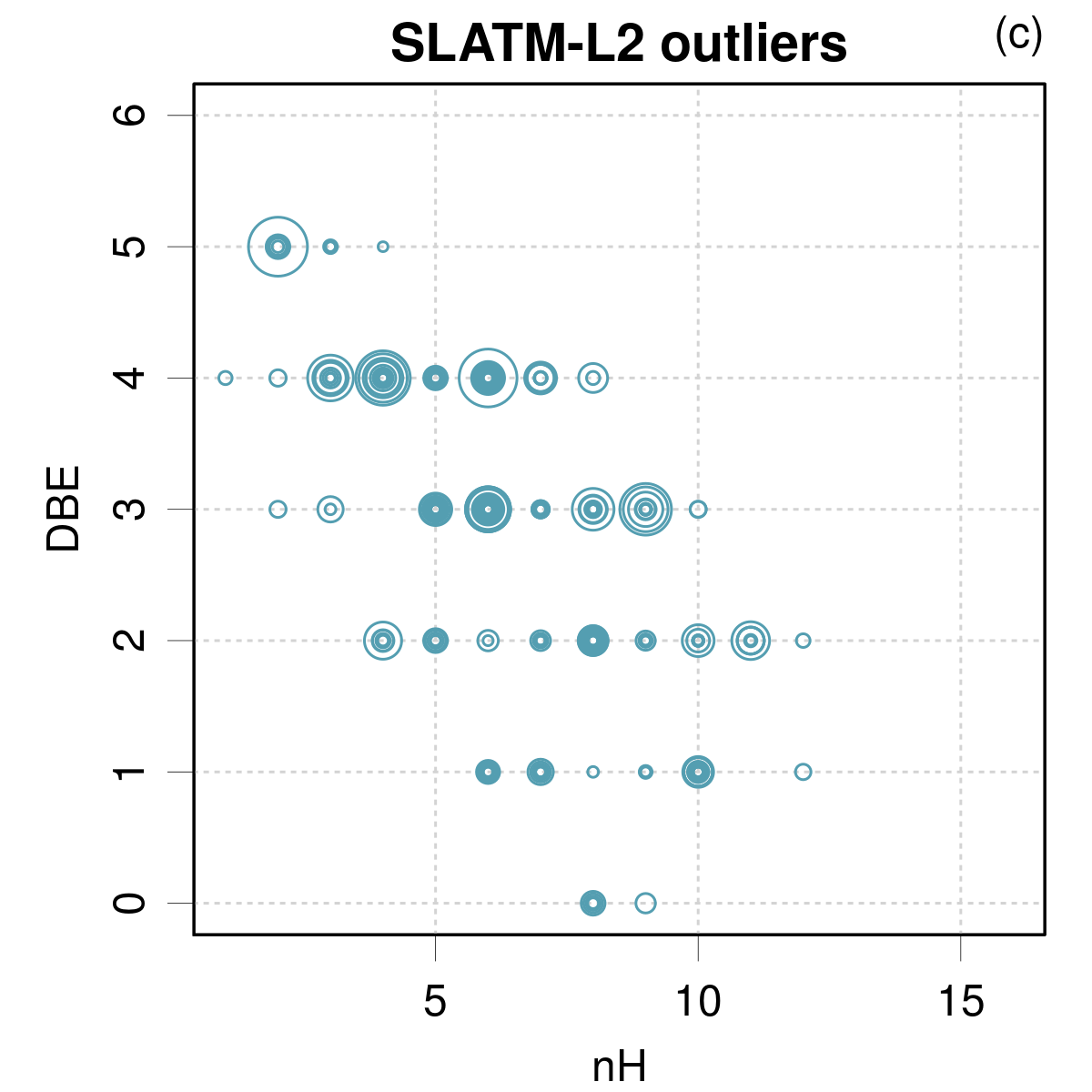}  & \includegraphics[width=0.45\textwidth]{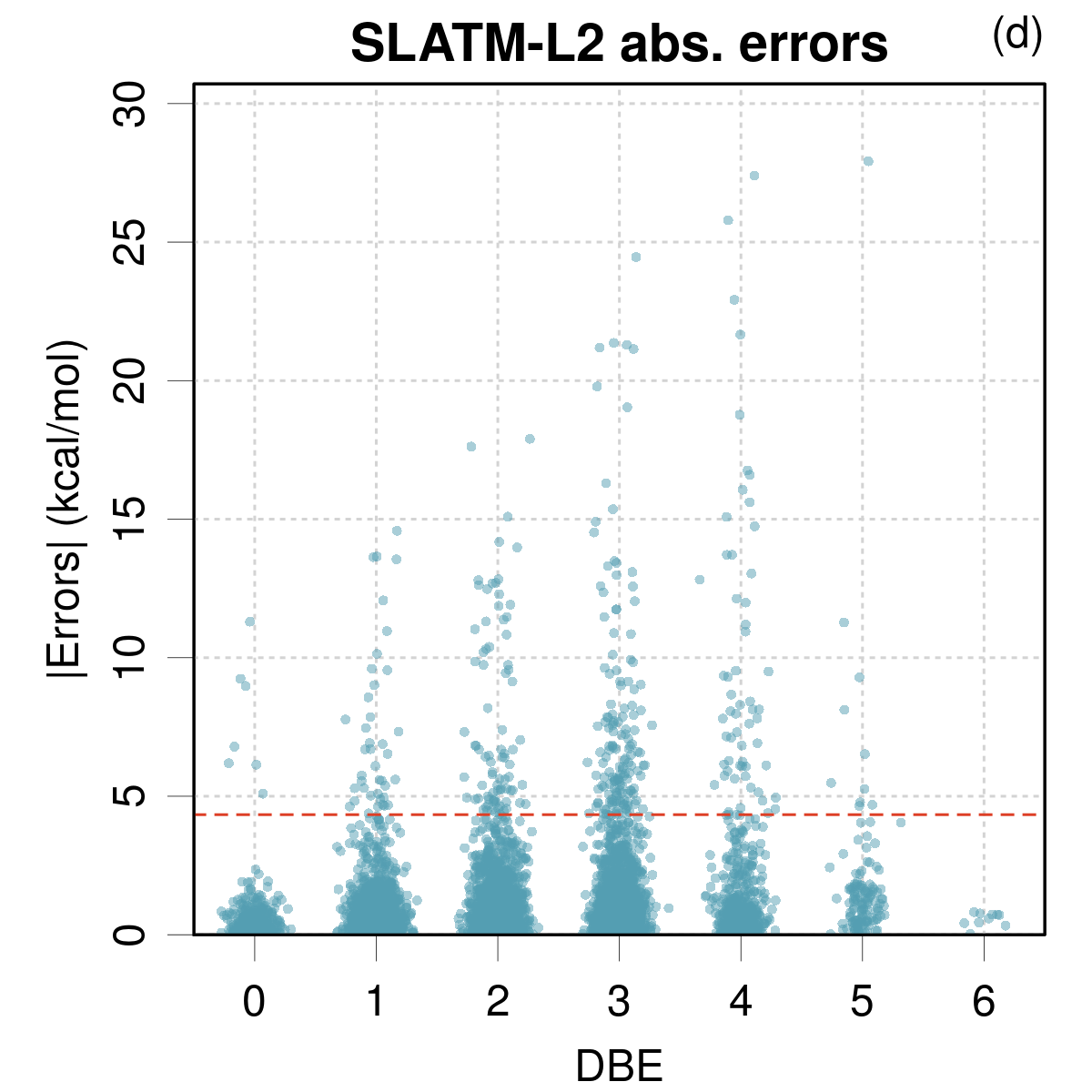} \tabularnewline
\end{tabular}\caption{\label{fig:ml3}Error distributions of SLATM-L2 (a) and MP2 (b) and
outliers of SLATM-L2 (c) as functions of the number of H atoms and
Double Bond Equivalent (DBE) of the molecules. The circles diameters
are on a common scale for all figures and are proportional to the
absolute errors. The distribution of SLATM-L2 errors for each DBE
class is shown in (d), where the horizontal dashed line represents
the $5*\sigma_{1}$ selection threshold for outliers.}
\end{figure}

The distribution of the absolute errors sorted by DBE class is shown
in Fig.\,\ref{fig:ml3}(d). Distributions of the absolute errors
according to $n\mathrm{H}$ and DBE are presented in (Fig.\,\ref{fig:ml3}(a,b)),
where each datum is represented by a circle of diameter proportional
to its absolute error. The scale is identical for all the series of
plots. The imprint of the points cloud reflects the correlation between
$n\mathrm{H}$ and DBE. 

For SLATM-L2, some systems with DBE\,=\,3-5 present very large errors,
and the maximal absolute error increases almost linearly from DBE\,=\,0
to 4. All compositions with DBE\,=\,6 are correctly predicted, although,
except for \ce{C5H2N2}, none of those compositions are found in the
learning set. For molecules with DBE\,=\,0, all heavy atoms are
sp$^{3}$-hybridized, being the most local and thus transferable and
easy to learn by QML models. For molecules with DBE\,=\,6, bond
environments are non-local, but could be easily covered by the training
set. Therefore, low prediction errors are expected in these cases. 

For MP2, the points cloud in Fig.\,\ref{fig:ml3}(b) presents no
strong feature, but the errors seem smaller for DBE\,=\,0, and for
each DBE class, there seems to be a maximum around the mid-range of
admissible $n\mathrm{H}$ values.

\subsection{Outliers analysis\label{subsec:Outliers-analysis}}

\begin{figure}[t]
\noindent \centering{}%
\begin{tabular}{cc}
\includegraphics[width=0.45\textwidth]{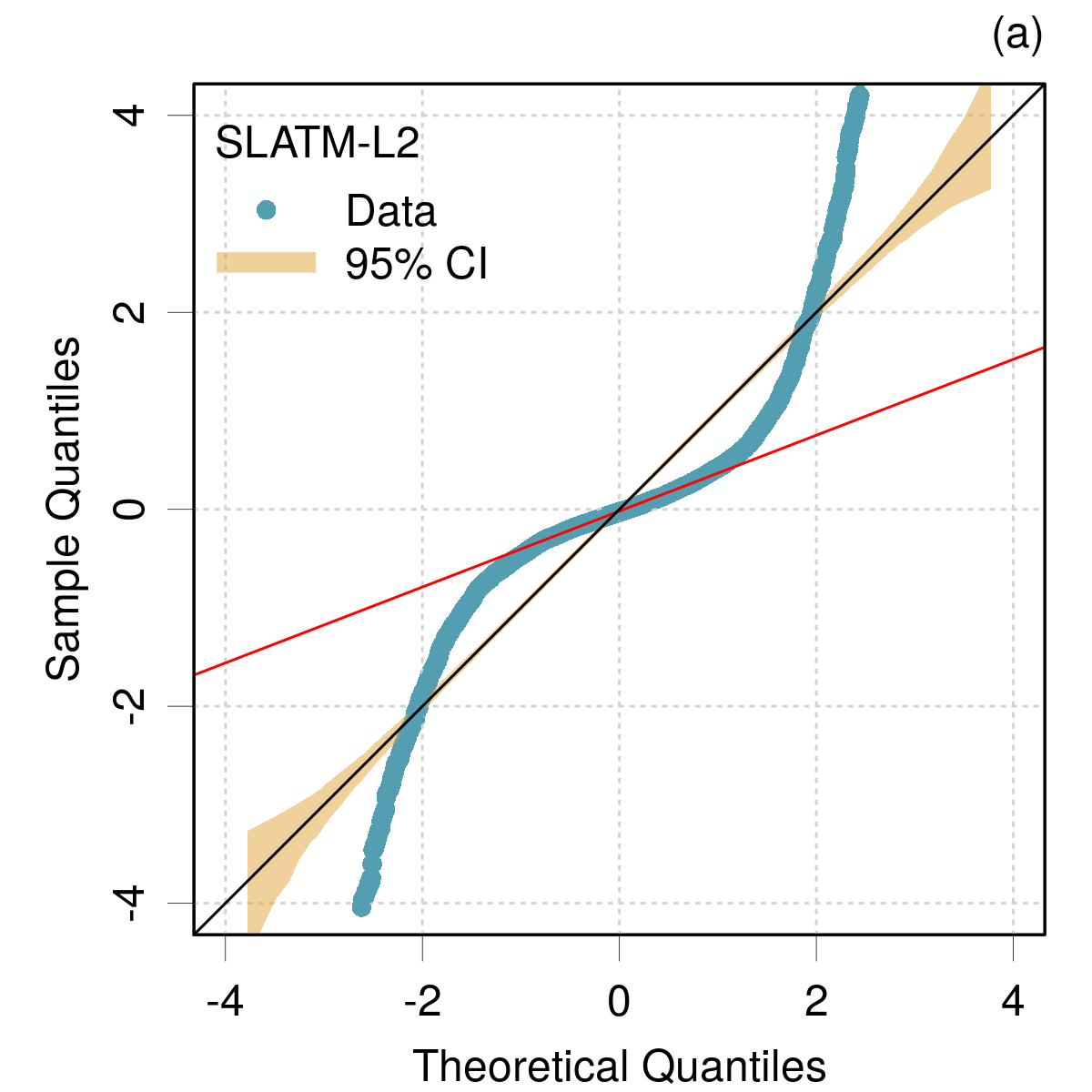}  & \includegraphics[width=0.45\textwidth]{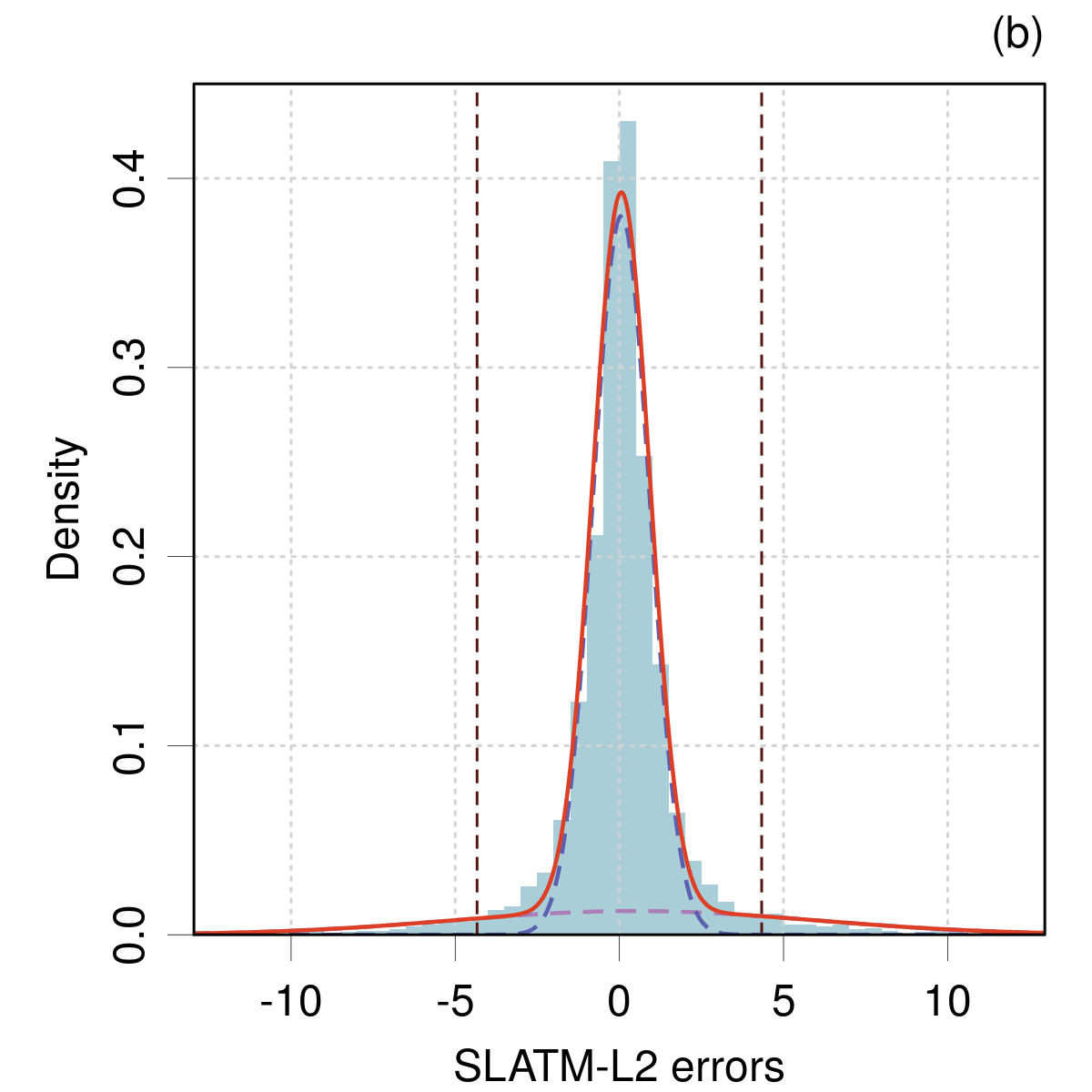}\tabularnewline
 & %
\begin{tabular}{cccc}
\multicolumn{4}{c}{\textbf{Parameters of the bi-normal fit} }\tabularnewline
\hline 
$i$  & $\mu_{i}$ (kcal/mol)  & $\sigma_{i}$ (kcal/mol)  & $w_{i}$\tabularnewline
\hline 
1  & -0.06  & 0.87  & 0.83 \tabularnewline
2  & -0.47  & 5.50  & 0.17 \tabularnewline
\hline 
\end{tabular}\tabularnewline
\end{tabular}\caption{\label{fig:ml4}Non-normality of the SLATM-L2 errors distribution.
(a) Normal QQ-plot: the orange area depicts the Monte Carlo estimated
statistical variability for a normal sample of the same size; the
red line passes through the first and third quartiles of the sample.
(b) Optimal bi-normal density superimposed on errors histogram: the
dashed vertical lines delimit the $5*\sigma_{1}$ threshold for outliers
identification.}
\end{figure}

The non-normality of the errors distribution for SLATM-L2 is clearly
visible on the QQ-plot in Fig.\,\ref{fig:ml4}(a). There is a near
linear section at the center if the distribution, seen by comparison
with the interquartile line, and also linear sections in the tails
of the distribution. The core of the distribution is therefore dominated
by a normal component, as well as the tails, albeit with different
widths (and centers). Indeed, a bi-normal fit (Eq.\,\ref{eq:binorfit})
provides a good representation of the distribution. Note that to get
a perfect fit, as much as four components are needed, but the bi-normal
representation is sufficient for the present purpose of outliers analysis.

The graphical results and optimal parameters are reported in Fig.\,\ref{fig:ml4}(b).
This decomposition provides a strong component (weight 0.83) with
a sub-kcal/mol standard deviation ($\sigma_{1}=0.87$\,kcal/mol),
and a weaker component (weight: 0.17) with a much larger dispersion
($\sigma_{2}=5.5$\,kcal/mol). Both components are nearly zero-centered.
 
\begin{table}[!th]
\begin{singlespace}
\noindent \begin{centering}
{\footnotesize{}}%
\begin{tabular}{lccccc}
\hline 
\textbf{\footnotesize{}Composition}{\footnotesize{} } & \textbf{\footnotesize{}DBE }{\footnotesize{} } & \textbf{\footnotesize{}medAE }{\footnotesize{} } & \textbf{\footnotesize{}maxAE }{\footnotesize{} } & \textbf{\footnotesize{}nOutl/nTest }{\footnotesize{} } & \textbf{\footnotesize{}nLearn }\tabularnewline
\hline 
{\footnotesize{}\ce{C4H2N2O} } & {\footnotesize{}5 } & {\footnotesize{}10.28 } & {\footnotesize{}27.92 } & {\footnotesize{}4/12 } & {\footnotesize{}3 }\tabularnewline
{\footnotesize{}\ce{C5H6N2} } & {\footnotesize{}4 } & {\footnotesize{}6.28 } & {\footnotesize{}27.40 } & {\footnotesize{}10/60 } & {\footnotesize{}17 }\tabularnewline
{\footnotesize{}\ce{C4H4N2O} } & {\footnotesize{}4 } & {\footnotesize{}8.13 } & {\footnotesize{}25.79 } & {\footnotesize{}14/64 } & {\footnotesize{}13 }\tabularnewline
{\footnotesize{}\ce{C6H9N} } & {\footnotesize{}3 } & {\footnotesize{}9.14 } & {\footnotesize{}24.46 } & {\footnotesize{}9/152 } & {\footnotesize{}25 }\tabularnewline
{\footnotesize{}\ce{C4H3NO2} } & {\footnotesize{}4 } & {\footnotesize{}10.93 } & {\footnotesize{}21.66 } & {\footnotesize{}4/25 } & {\footnotesize{}3 }\tabularnewline
{\footnotesize{}\ce{C4H6N2O} } & {\footnotesize{}3 } & {\footnotesize{}6.22 } & {\footnotesize{}21.36 } & {\footnotesize{}37/223 } & {\footnotesize{}39 }\tabularnewline
{\footnotesize{}\ce{C5H6O2} } & {\footnotesize{}3 } & {\footnotesize{}6.07 } & {\footnotesize{}21.29 } & {\footnotesize{}11/115 } & {\footnotesize{}22 }\tabularnewline
{\footnotesize{}\ce{C6H8O} } & {\footnotesize{}3 } & {\footnotesize{}6.02 } & {\footnotesize{}19.79 } & {\footnotesize{}10/203 } & {\footnotesize{}33 }\tabularnewline
{\footnotesize{}\ce{C6H11N} } & {\footnotesize{}2 } & {\footnotesize{}12.71 } & {\footnotesize{}17.90 } & {\footnotesize{}6/207 } & {\footnotesize{}29 }\tabularnewline
{\footnotesize{}\ce{C3H4O3S} } & {\footnotesize{}2 } & {\footnotesize{}9.86 } & {\footnotesize{}17.62 } & {\footnotesize{}5/6 } & {\footnotesize{}1 }\tabularnewline
{\footnotesize{}\ce{C4H3NOS} } & {\footnotesize{}4 } & {\footnotesize{}13.13 } & {\footnotesize{}16.76 } & {\footnotesize{}2/6 } & {\footnotesize{}2 }\tabularnewline
{\footnotesize{}\ce{C5H4OS} } & {\footnotesize{}4 } & {\footnotesize{}16.60 } & {\footnotesize{}16.60 } & {\footnotesize{}1/3 } & {\footnotesize{}0 }\tabularnewline
{\footnotesize{}\ce{C4H5NO2} } & {\footnotesize{}3 } & {\footnotesize{}8.44 } & {\footnotesize{}15.36 } & {\footnotesize{}20/109 } & {\footnotesize{}18 }\tabularnewline
{\footnotesize{}\ce{C6H10O} } & {\footnotesize{}2 } & {\footnotesize{}5.32 } & {\footnotesize{}15.09 } & {\footnotesize{}4/242 } & {\footnotesize{}51 }\tabularnewline
{\footnotesize{}\ce{C6H7N} } & {\footnotesize{}4 } & {\footnotesize{}9.48 } & {\footnotesize{}15.08 } & {\footnotesize{}6/51 } & {\footnotesize{}6 }\tabularnewline
{\footnotesize{}\ce{C4H6N2S} } & {\footnotesize{}3 } & {\footnotesize{}6.71 } & {\footnotesize{}14.91 } & {\footnotesize{}14/40 } & {\footnotesize{}8 }\tabularnewline
{\footnotesize{}\ce{C4H10N2S} } & {\footnotesize{}1 } & {\footnotesize{}13.63 } & {\footnotesize{}14.58 } & {\footnotesize{}5/5 } & {\footnotesize{}0 }\tabularnewline
{\footnotesize{}\ce{C4H8N2S} } & {\footnotesize{}2 } & {\footnotesize{}12.10 } & {\footnotesize{}14.18 } & {\footnotesize{}12/12 } & {\footnotesize{}0 }\tabularnewline
{\footnotesize{}\ce{C7H8} } & {\footnotesize{}4 } & {\footnotesize{}9.98 } & {\footnotesize{}13.71 } & {\footnotesize{}2/43 } & {\footnotesize{}7 }\tabularnewline
{\footnotesize{}\ce{C6H6O} } & {\footnotesize{}4 } & {\footnotesize{}8.67 } & {\footnotesize{}13.71 } & {\footnotesize{}5/66 } & {\footnotesize{}6 }\tabularnewline
{\footnotesize{}\ce{C3H7N1O2S} } & {\footnotesize{}1 } & {\footnotesize{}6.88 } & {\footnotesize{}12.07 } & {\footnotesize{}9/15 } & {\footnotesize{}1 }\tabularnewline
{\footnotesize{}\ce{C3H3NO2S} } & {\footnotesize{}3 } & {\footnotesize{}8.57 } & {\footnotesize{}12.04 } & {\footnotesize{}2/2 } & {\footnotesize{}1 }\tabularnewline
{\footnotesize{}\ce{C6H6} } & {\footnotesize{}4 } & {\footnotesize{}11.99 } & {\footnotesize{}11.99 } & {\footnotesize{}1/5 } & {\footnotesize{}5 }\tabularnewline
{\footnotesize{}\ce{C5H8O2} } & {\footnotesize{}2 } & {\footnotesize{}5.72 } & {\footnotesize{}11.87 } & {\footnotesize{}6/229 } & {\footnotesize{}26 }\tabularnewline
{\footnotesize{}\ce{C3H5NO2S} } & {\footnotesize{}2 } & {\footnotesize{}7.39 } & {\footnotesize{}11.31 } & {\footnotesize{}5/5 } & {\footnotesize{}2 }\tabularnewline
{\footnotesize{}\ce{C3H8O3S} } & {\footnotesize{}0 } & {\footnotesize{}6.49 } & {\footnotesize{}11.30 } & {\footnotesize{}6/8 } & {\footnotesize{}0 }\tabularnewline
{\footnotesize{}\ce{C5H5NO} } & {\footnotesize{}4 } & {\footnotesize{}6.08 } & {\footnotesize{}11.20 } & {\footnotesize{}7/83 } & {\footnotesize{}10 }\tabularnewline
{\footnotesize{}\ce{C3H6O3S} } & {\footnotesize{}1 } & {\footnotesize{}7.77 } & {\footnotesize{}10.96 } & {\footnotesize{}9/14 } & {\footnotesize{}4 }\tabularnewline
{\footnotesize{}\ce{C5H10N2} } & {\footnotesize{}2 } & {\footnotesize{}5.21 } & {\footnotesize{}10.83 } & {\footnotesize{}4/194 } & {\footnotesize{}28 }\tabularnewline
{\footnotesize{}\ce{C4H8N2O} } & {\footnotesize{}2 } & {\footnotesize{}5.66 } & {\footnotesize{}10.20 } & {\footnotesize{}20/245 } & {\footnotesize{}50 }\tabularnewline
{\footnotesize{}\ce{C4H10N2O} } & {\footnotesize{}1 } & {\footnotesize{}5.56 } & {\footnotesize{}10.14 } & {\footnotesize{}9/151 } & {\footnotesize{}25 }\tabularnewline
\hline 
\end{tabular}{\footnotesize\par}
\par\end{centering}
\end{singlespace}
\begin{centering}
 
\par\end{centering}
\noindent \caption{\label{tab:ML5soutl}Composition of the 5$\sigma$ outliers with maximum
absolute error above 10 kcal/mol. For each composition, one defines
medAE: median Absolute error in outlier isomers; maxAE: max Absolute
error in outlier isomers; nOutl: number of isomers in the outlier
set; nTest: number of isomers in the test set; nLearn: number of isomers
in the learning set.}
\end{table}

In order to identify the systems involved in the widest contribution,
one selects outliers with respect to the most concentrated component,
\emph{i.e.} points lying at more than $5*\sigma_{1}$ from the center
of the concentrated component ($\mu_{1}$). 

This filter selects 350 molecules (about 6\,\% of the test set population).
The chemical formulae of these systems are analyzed in terms of $n\mathrm{H}$
and DBE (Fig.~\ref{fig:ml3}(c)). Systems with DBE\,=\,0 or 6 are
practically absent from the list of outliers (at the exception of
a few DBE\,=\,0 ones). The outliers with the largest absolute errors
are molecules with some unsaturation (DBE $\ge$ 3). 
\begin{figure}[!t]
\noindent \centering{}%
\begin{tabular}{cc}
\includegraphics[width=0.35\textwidth]{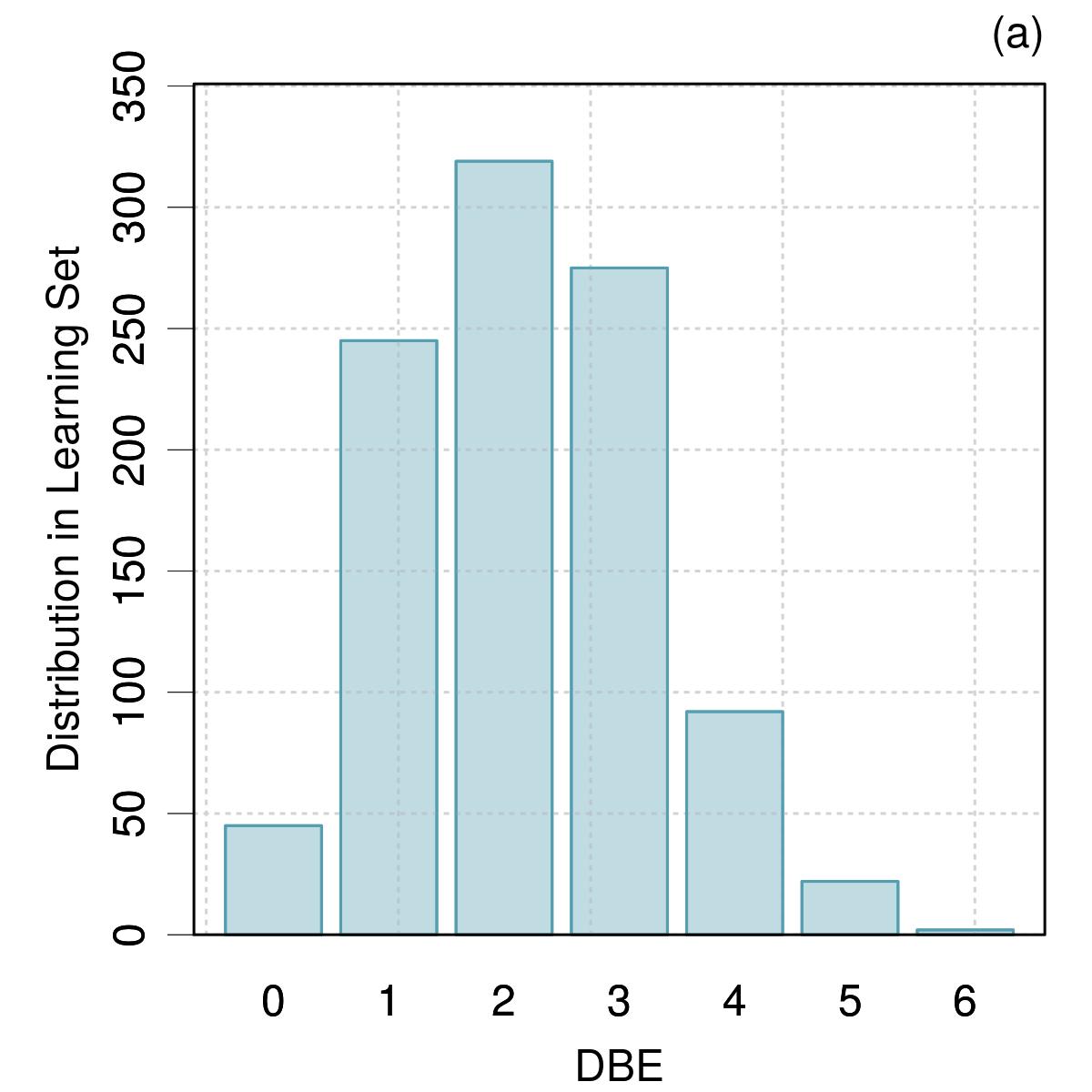}  & \includegraphics[width=0.35\textwidth]{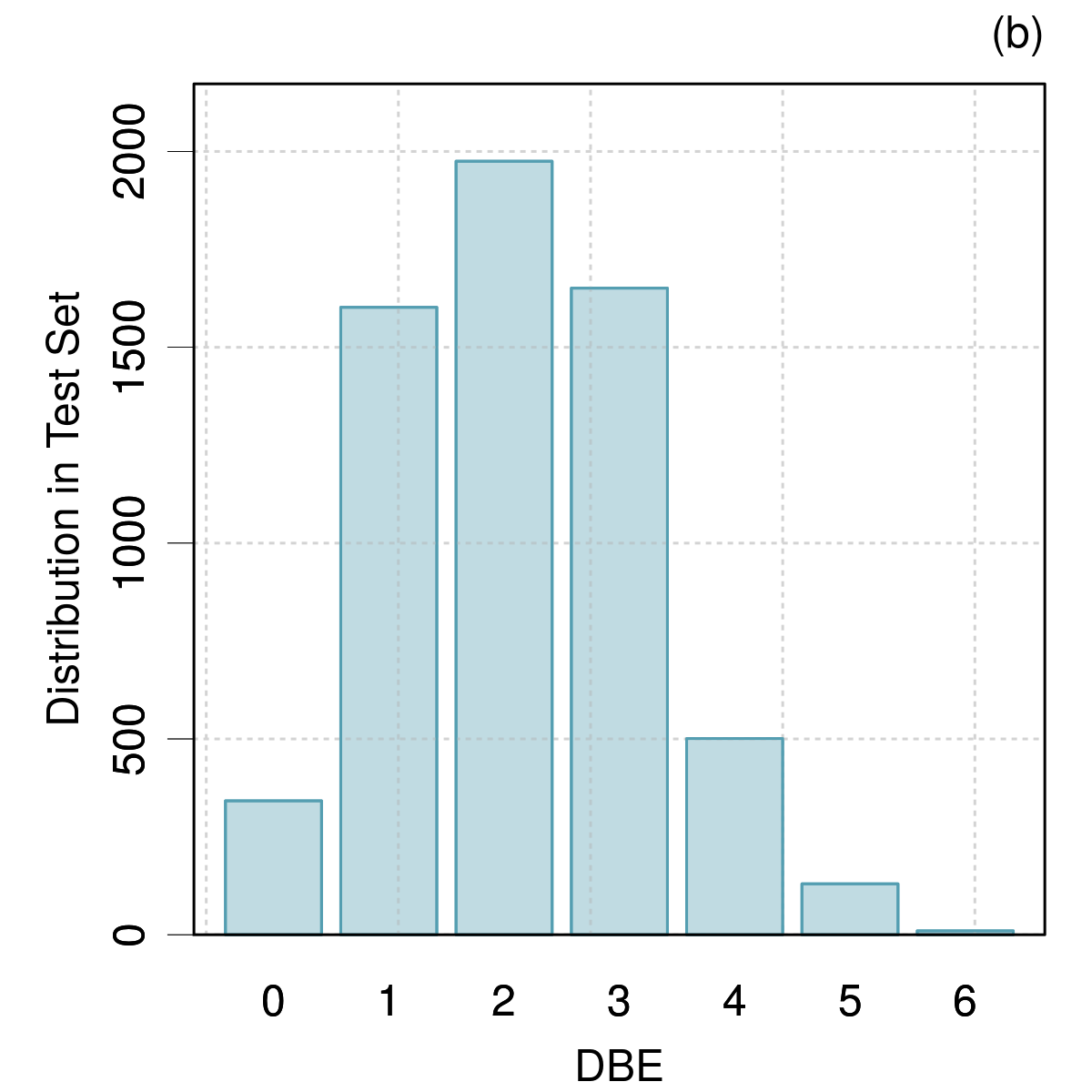} \tabularnewline
\includegraphics[width=0.35\textwidth]{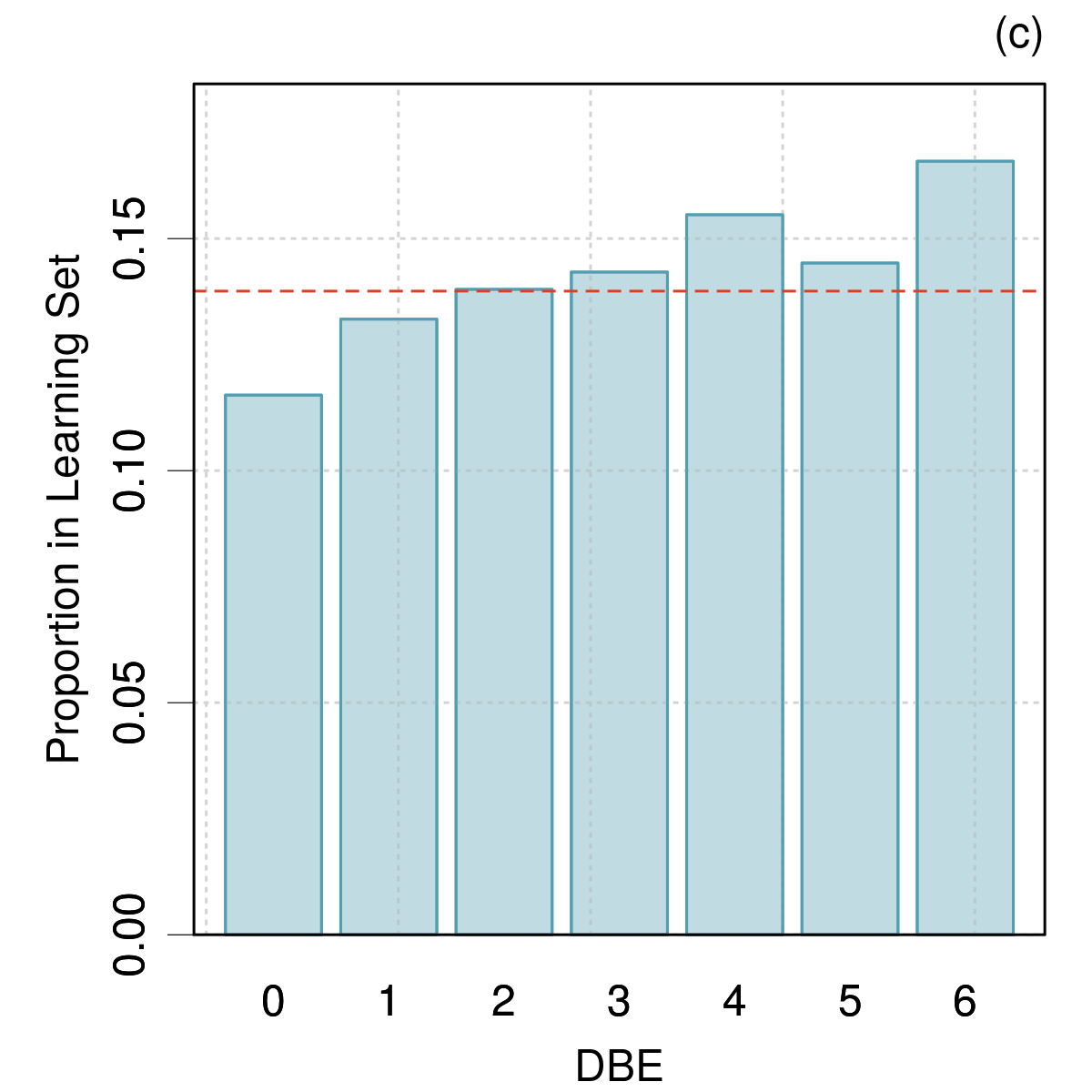}  & \includegraphics[width=0.35\textwidth]{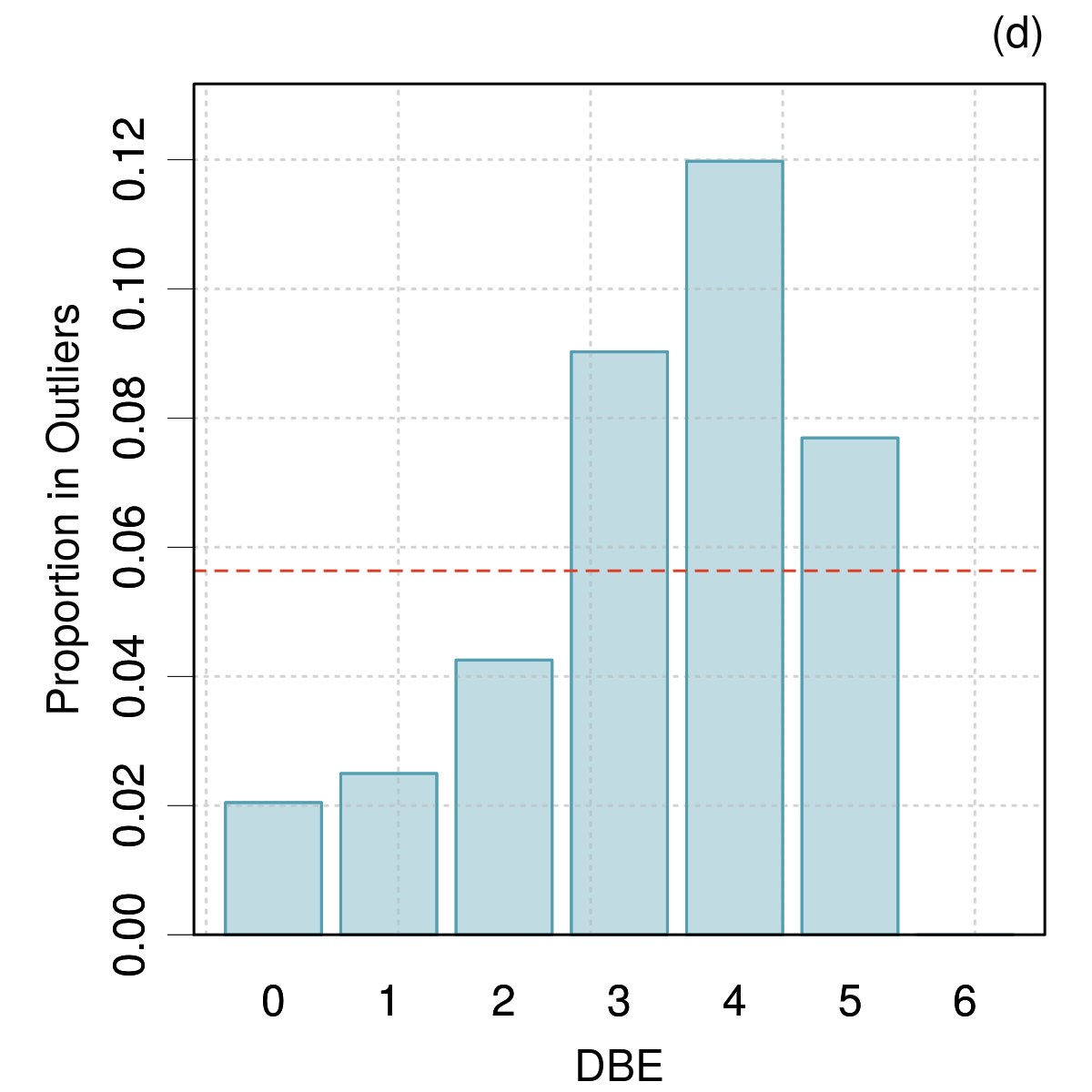} \tabularnewline
\end{tabular}\caption{\label{fig:mla1}DBE distribution in (a) the learning set and (b)
the test set; (c) proportion of each DBE class in the learning set
with respect to the distribution in the full data set (learning +
test); (d) proportion of SLATM-L2 outliers in each DBE class with
respect to the distribution in the test set. The horizontal dashed
lines represent the mean proportion over the reference dataset.}
\end{figure}

Fig.~\ref{fig:mla1} reports the DBE distributions in the leaning
set Fig.~\ref{fig:mla1}(a) and test set Fig.~\ref{fig:mla1}(b).
In both cases, the distributions are peaked around DBE=2. The distributions
are similar because the learning set is a random selection in the
full dataset. The ratio of the learning set DBE distribution over
the full dataset DBE distribution is practically constant, with slight
increasing trend that might be a random effect (Fig.~\ref{fig:mla1}(c)).
In contrast, Fig.~\ref{fig:mla1}(d) shows that the DBE classes around
DBE=4 contain a larger proportion of outliers than the lower DBE classes,
up to 12\,\% for DBE\,=\,4. Outliers are not uniformly distributed
amongst the DBE classes.

Compositions with a maximum absolute error larger than 10\,kcal/mol
are listed in Table~\ref{tab:ML5soutl}. The composition with the
largest absolute error is \ce{C4H2N2O} (27.9\,kcal/mol for the molecule
SMILES:C\#CC(=N\#N)C=O \citep{Weininger1988}). This molecule, with
DBE=5, has 3 isomers/configurations in the learning set and 12 in
the test set, 4 of which are outliers. The second one is \ce{C5H6N2}
(27.4\,kcal/mol, SMILES:c1cnn2c1CC2), with 17 configurations in the
learning set, 60 in the test set, 10 of which are outliers.

One can see in this list that most outliers compositions present some
isomers in the learning set. Unsurprisingly, it contains also a few
compositions absent from the learning set, \emph{e.g.}, \ce{C5H4OS}
(SMILES:c1c2c(cs1)OC2), but as seen above for the DBE=6 compounds,
this is not systematically associated with poor predictions.

The impact on the error statistics of the removal of the 350 outliers
from the test set is reported in Table\,\ref{tab:MLStats}, as method
SLATM-L2(-o). The improvement is noticeable on all aspects: better
normality of the distribution, reduced bias (MSE\,$\simeq0$), and
a notable reduction, as expected, of MAE, RMSD and $Q_{95}$.

\subsection{Expanding the learning set with outliers\label{subsec:Expanding-the-learning}}

Having identified 350 systems for which SLATM-L2 provides exceedingly
large errors, we want to check if their transfer to the learning set
has a positive impact on prediction performances. To validate the
impact of this learning set augmentation, a list of 350 systems has
been transferred from the test set to the learning set according to
two selections: (1) the outliers identified above (SLATM-L2(+o)),
and (2) a random choice (SLATM-L2(+r)), both resulting in a learning
set of size 1350. The second scenario is a control to assess the size
effect of the learning set. Note that the present random selection
contains 22 of the 350 outliers ($\simeq$\,6\,\%), in proportion
with the random selection of 350 points among 6211.

\begin{figure}[!t]
\noindent \centering{}%
\begin{tabular}{cc}
\includegraphics[width=0.27\textheight]{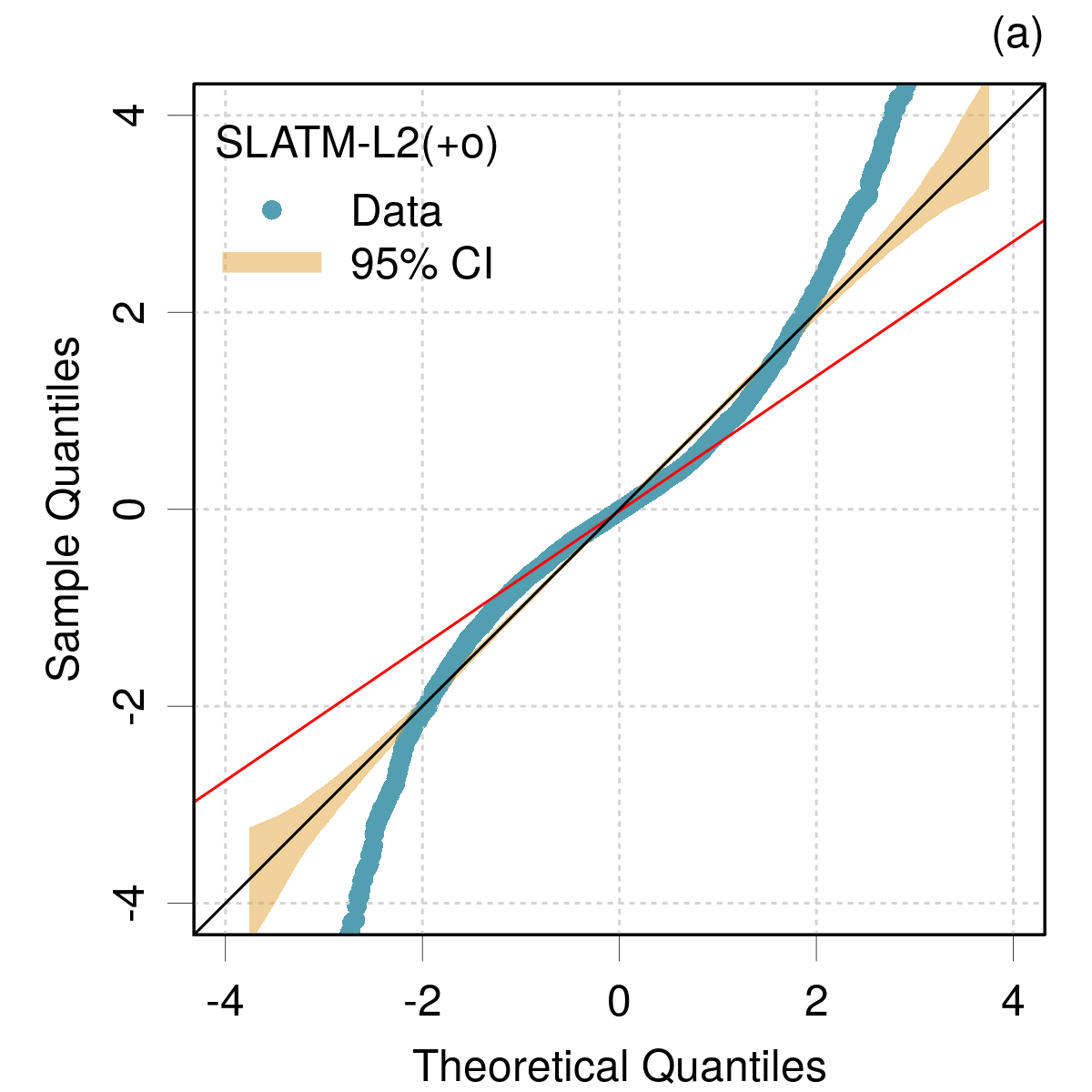} & \includegraphics[width=0.27\textheight]{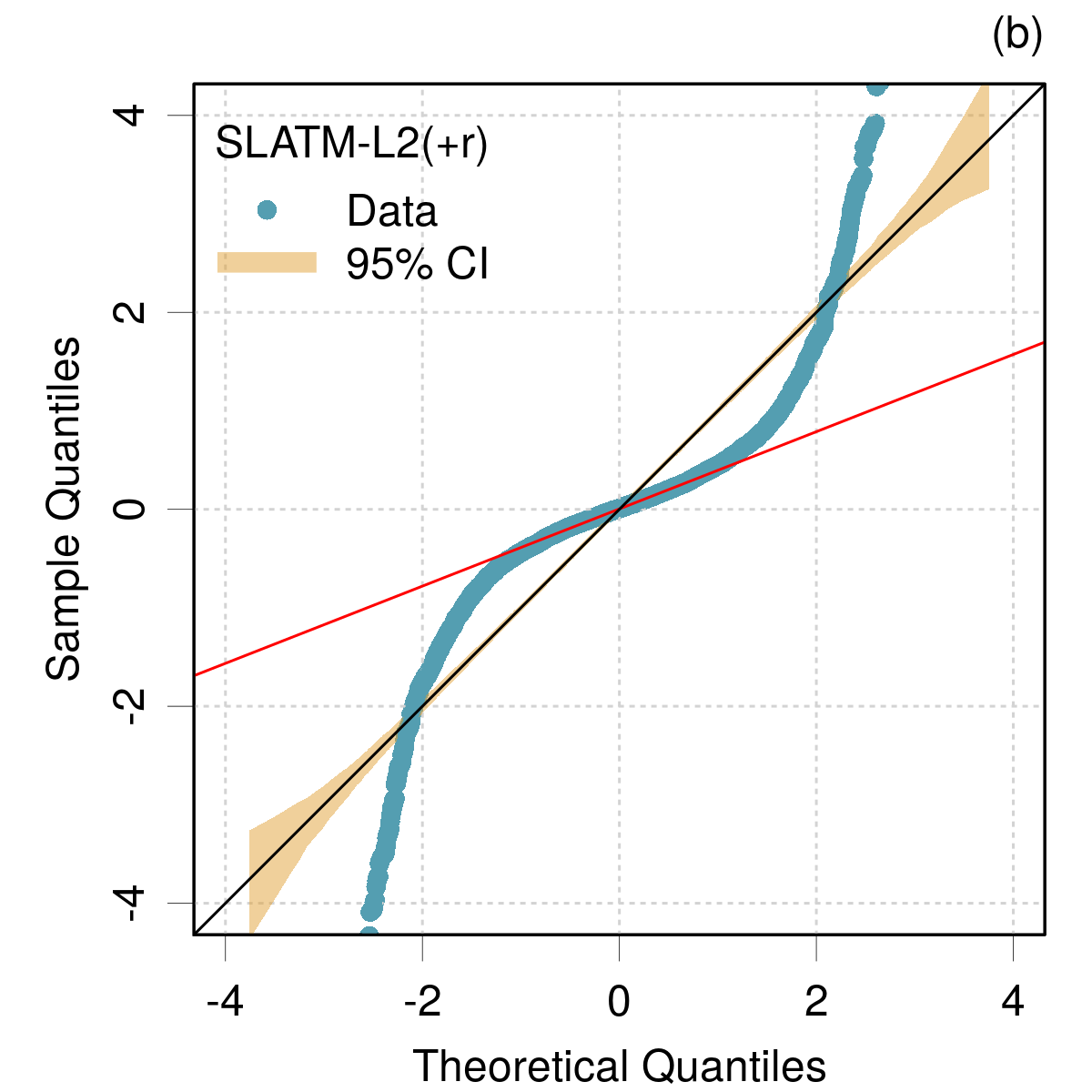}\tabularnewline
\includegraphics[width=0.27\textheight]{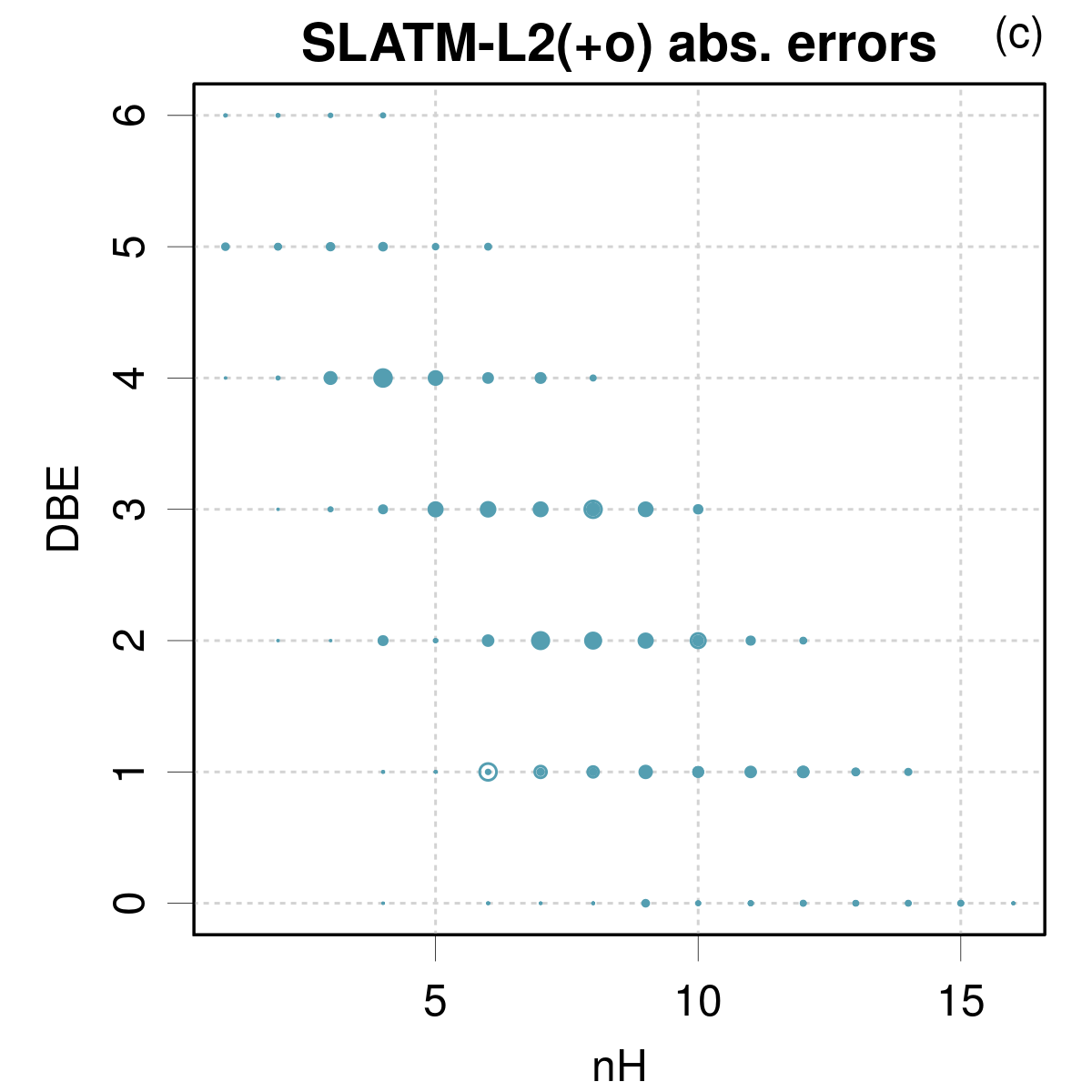}  & \includegraphics[width=0.27\textheight]{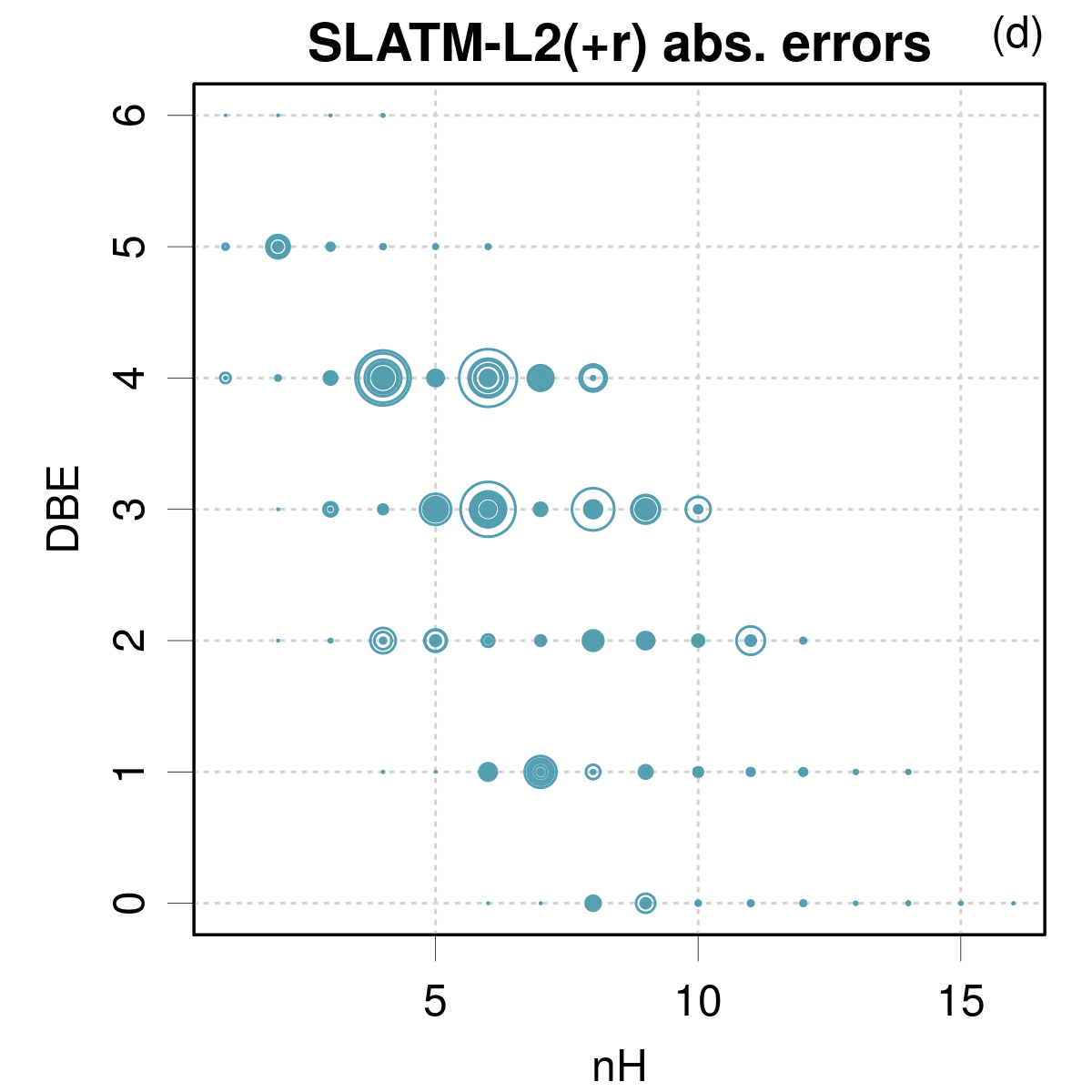}\tabularnewline
\end{tabular}\caption{\label{fig:ml6}Normal QQ-plots of the error distributions for (a)
SLATM-L2(+o) and (b) SLATM-L2(+r) (see Fig.\,\ref{fig:ml4} for details);
(c,d): absolute error distributions as functions of the number of
H atoms and Double Bond Equivalent (DBE) of the molecules. The circles
diameters are on a common scale for all figures and are proportional
to the absolute errors.}
\end{figure}

The SLATM-L2 model has been retrained with both learning sets, and
the error distribution analysis has been performed for these two new
datasets. The results are reported in Table\,\ref{tab:MLStats} and
Fig.\,\ref{fig:ml1}(e,f), \ref{fig:ml1-1}(b), Fig.\,\ref{fig:ml6}.

A first aspect to consider is how the inclusion of the outliers in
the learning set (SLATM-L2(+o)) affects the prediction errors with
respect to the pruned error set (SLATM-L2(-o)) presented in the previous
section. Globally, the improvement is minor, but consistent. The MSE
and RMSD are identical, the MAE is barely decreased, and the $Q_{95}$
is slightly improved from 2.70 to 2.49\,kcal/mol. The learning on
the outliers has been transferred to some of the predictions.

Then, we compare the gain in prediction quality due to the augmentation
of the learning set by outliers versus random systems. The MAE values
of SLATM-L2(+o):\,0.83 and SLATM-L2(+r):\,0.95\,kcal/mol indicate
a statistically significant, but rather weak effect of the outliers
selection, whereas the sheer augmentation of the learning set size
improves notably the MAE of the original method by 0.3\,kcal/mol.
The MSE shows that the slight bias of the original SLATM-L2 has been
essentially corrected in both cases. Impact on the RMSD is larger
than on the MAE (1.20 for '+o' \emph{vs}. 1.89\,kcal/mol for '+r'),
with a notable reduction of 1.24\,kcal/mol from the initial RMSD.
Similarly, the decrease in $Q_{95}$ is important for SLATM-L2(+o),
from 4.7 to 2.49\,kcal/mol, and much less for SLATM-L2(+r), to 3.15\,kcal/mol,
which reaches the level of MP2 (3.34\,kcal/mol). One sees also that
the normality of the distribution has been improved for SLATM-L2(+o)
(Kurt from 29 to 8.1; $W$ from 0.71 to 0.935), which is not the case
for the random selection (Kurt =46; $W$\,=\,0.66).

Without looking at the plots of the distributions, this set of statistics
points to a notable improvement of the error distribution on all criteria
for the outliers-augmented learning set. By contrast, the improvement
by random selection of the 350 systems, although significant, does
not improve the quality of the error distribution, which stays strongly
leptokurtic. This can be appreciated on the QQ-plots for both error
distributions in Fig.\,\ref{fig:ml6}(a,b). 

The ECDFs of the error distributions illustrate clearly the analysis
of the summary statistics (Fig.\,\ref{fig:ml1-1}(b)). The augmented
SLATM-L2 versions have more concentrated absolute errors than SLATM-L2.
In the case of SLATM-L2(+o) the large errors have essentially vanished
and the overall performance is better than SLATM-L2 and even than
MP2. Albeit SLATM-L2(+r) is also more concentrated, it still presents
a set of systems with large errors, exceeding those of MP2. The comparison
of the $n\mathrm{H}$/DBE plots in Fig.\,\ref{fig:ml6}(c,d) and
depicts precisely this point.

\section{Discussion\label{sec:Discussion}}

\noindent We analyzed the prediction error distribution of a ML method
(KRR-SLATM-L2) for effective atomization energies of QM7b molecules
calculated at the level of theory CCSD(T)/cc-pVDZ. Error distributions
of standard computational chemistry methods (HF and MP2) at the same
basis set level and for the same reference dataset were also estimated
for comparison.

We have shown that the shapes of error distributions should carefully
be considered when one attempts to quantify the prediction performance
of a method. MAE-based benchmarks neglect crucial information, that
is better rendered by probabilistic estimators such as $Q_{95}$.
Similar values of the MAE might hide very different error distributions,
with a strong impact on the assessment and ranking of methods. In
particular, ML prediction error distributions can be strongly non-normal,
notably leptokurtic, which poses a problem of prediction reliability,
as they might have a non-negligible probability to produce very large
errors. Identification of systems with large prediction errors and
their inclusion into the learning set improves significantly the prediction
error distribution. These main points are further discussed below.

\subsection{Limitations of the present study}

The QM7b dataset has a nice feature of diversity, albeit it is small
in total size. A natural yet undesired consequence is that the dataset
is highly inhomogeneous in chemical space, the conclusions drawn above
from the results on QM7b may therefore not be extensible to other
datasets, such as QM9 \citep{Ruddigkeit2012,Ramakrishnan2014} (less
diverse in composition, yet larger in total set size). Considering
these properties of QM9, one might expect the prediction error distribution
for SLATM-L2 to be less leptokurtic than for QM7b, and a less problematic
use of MAE for performance assesment. This has however to be checked,
but at the moment, one has no access to hierarchical properties for
the full QM9, as for QM7b.

In this paper we assessed the prediction of energy, which is extensive
in nature. There exists a set of intensive properties as well and
which are typically difficult to learn. The generalizability of our
observations to these properties should also be confirmed in future
studies. 

\subsection{Analysis of the SLATM-L2 error distribution}

A major feature of the SLATM-L2 error distribution is the presence
of a strong peak of small errors and wide tails of large errors. This
leptokurtic distribution has been successfully modeled as a mixture
of two normal distributions, both practically centered on zero, with
standard deviations 0.87 and 5.5\,kcal/mol and a mixture ratio 83:17.
Based on this analysis, we explore the chemical identity of the systems
with large errors. 

It is not possible to unscramble both populations in the area where
the most concentrated population is dominant, but if one goes far
enough in the tails (beyond 5 times 0.87\,kcal/mol) one finds a large
majority of systems belonging to the population with large errors.
Such 350 systems were identified and tagged as outliers. A chemical
analysis of these systems based on their DBE revealed that compositions
with high levels of unsaturation (DBE$=3-5$) are overrepresented
with respect to their abundance in the test set. It seems therefore
that the large prediction errors are not randomly distributed over
the QM7b molecules, which might provide insights for improved molecular
descriptors or a better design of the learning set.

In fact, if the learning set is augmented with these outliers (SLATM-L2(+o)),
one gets better prediction performances than if one augments the learning
set with 350 randomly chosen systems (SLATM-L2(+r)). This suggests
an iterative method for the design of learning sets that might deserve
further consideration.

\subsection{Importance for ML to aim at near-normality of error distributions }

For ML methods to be used trustfully as replacements of quantum chemistry
methods, one needs to have a reliable measure of their accuracy. Ideally,
this should be quantified by a single number characterizing the predictive
ability of a method. This requires a finite variance, homogeneous,
error distribution (with a dispersion that does not vary strongly
within the calculated property range) and a zero-centered symmetric
error distribution. Moreover, we have seen that leptokurtic error
distributions with a strong core of small errors and tails of large
errors, as described above by a bi-normal distribution might not be
properly summarized by a single dispersion statistic. The MAE is not
a good choice in such cases, and $Q_{95}$ presents a more reliable
alternative, which quantifies the risk of large prediction errors,
even in the case of non-zero-centered symmetric error distributions.

We have shown that inclusion of outliers in the learning set of the
SLATM-L2 method could result in a notable improvement of the shape
of its prediction errors distribution. One might also presume that
the use of much larger learning sets should improve the structure
of the prediction error distribution.

\subsection{What probability does one have to reach the chemical accuracy ?}

Whatever the pertinence of this question, it cannot be answered by
considering the MAE, contrary to a still too common practice in the
benchmarking literature to use it as an ``accuracy'' measure. The
example of the SLATM-L2(+o) method shows that a sub-chemical-accuracy
MAE (0.86\,kcal/mol) can hide a non-negligible risk (about 30\,\%)
to exceed this limit. 

A direct answer to the question is provided by the ECDF of the errors
and $C(\eta)$ with $\eta=$1\,kcal/mol. The values reported in Table\,\ref{tab:MLStats}
tell us that none of the studied methods achieves the 1\,kcal/mol
threshold with a high probability. The best performers would be SLATM-L2
methods with augmented learning sets, with $C(1)=0.743$ for the (+r)
version. Even in this case, one has no firm guarantee to reach chemical
accuracy for any new prediction, the risk being higher than 25\,\%
to overpass the threshold. Note that this is a notable improvement
over MP2, for which the risk is about 55\,\%. Besides, this qualifies
chemical accuracy with respect to CCSD(T), not with respect to experimental
values.

\subsection{On the use of statistics to rank methods}

In Section\,\ref{subsec:Error-statistics}, we outlined that the
basic summary statistics (MAE, RMSD) provided contradictory information
about the SLATM-L2 prediction performances, notably when compared
to MP2. SLATM-L2 has a smaller slightly smaller MAE than MP2 (1.26
\emph{vs.} 1.31\,kcal/mol) but a much larger RMSD (2.44 \emph{vs.}
1.67). This is a consequence of the peculiar shape of the error distribution
for SLATM-L2. In line with the RMSD values, MP2 achieves a notably
smaller $Q_{95}$ value (3.34 vs. 4.7\,kcal/mol), meaning that SLATM-L2
is susceptible of larger errors than MP2. 

Given this set of information and computation time aside, one would
most certainly pick MP2 over SLATM-L2 as a reasonable predictor of
CCSD(T) energies, with a prediction uncertainty of approximately 1.7\,kcal/mol
for $E^{*}$ (RMSD in Table\,\ref{tab:MLStats}), and with added
confidence from the near-normality of the distribution, reducing the
risk of rogue predictions that might be expected from SLATM-L2. Finally,
the modified SLATM-L2(+o) method, for which the learning set was augmented
with SLATM-L2 outliers, clearly outperforms both SLATM-L2 and MP2
for all statistics. 

This illustrates once more that the MAE is not sensitive to major
differences in the shape of the errors distributions, as was shown
by Pernot and Savin \citep{Pernot2018}: even for normal distributions,
many combinations of MSE and RMSD can produce the same MAE. The case
is even worse for non-normal distributions, where the MAE loses its
ability to quantify prediction uncertainty. It is clear that the MAE
should not be used as a unique scoring statistic, and that it should
at least be complemented (or replaced) by probabilistic indicators
such as $Q_{95}$ .

\section*{Acknowledgments}

The authors are grateful to Anatole von Lilienfeld for fruitful discussions
and for his help with assembling the dataset.

\bibliographystyle{unsrturlPP}
\bibliography{NN,packages}

\end{document}